# Fracture mechanical behavior of polymers: 1. Amorphous glassy state

Travis Smith, Chaitanya Gupta, Caleb Carr, Shi-Qing Wang[*]

*School of Polymer Science and Polymer Engineering, University of Akron, OH 44325, USA*

*Abstract*

Theoretical analyses and experiments have been carried out to investigate fracture and failure behavior of glassy polymers, aiming to obtain new insights into the extreme mechanics of plastics. Our birefringence measurements quantify the local stress buildup at cut tip during different stages of drawing of a precut specimen. Based on brittle polymethyl methacrylate (PMMA), ductile bisphenol A polycarbonate (PC) and polyethylene terephthalate (PET), we find several key results beyond the existing knowledge base. (1) The inherent fracture and yield strengths $\sigma_{F(inh)}$ and $\sigma_{Y(inh)}$ differ little in magnitude from the breaking and yield stress ($\sigma_b$ and $\sigma_y$) respectively measured from uncut specimens. (2) Stress intensification (SI) near a pre-through-cut build up ceases because of finite tip sharpness. (3) The stress $\sigma_{tip}$ at cut tip shows a trend of approximate linear increase with the stress intensity factor $K_I = \sigma_0(\pi a)^{1/2}$ or far-field load $\sigma_0$ for all three polymers and different cut size $a$. (4) A characteristic length scale P emerges from the linear relation between $\sigma_{tip}$ and $K_I$. For these glassy polymers, P is on the order of 0.1 mm, apparently determined by the tip bluntness that occurs during the precut making. (5) Fracture toughness of brittle polymers is characterized by critical stress intensity factor $K_{Ic} = \sigma_{F(inh)}(2\pi P)^{1/2}$, revealing relevance of the two crucial quantities. (6) The critical energy release rate $G_{Ic}$ for brittle glass polymers such as PMMA is determined by the product of its work of fracture $w_F$ (of uncut specimen) and P. (7) The elusive fractocohesive length $L_{fc}$ defined in the literature as $G_{Ic}/w_F$ naturally arises from the new expression for $G_{Ic}$ as stated in (6), i.e., it is essentially proportional to P. These results suggest that a great deal of future work is required to acquire additional understanding with regards to fracture and failure behaviors of plastics.

## 1. Introduction

Fracture mechanics is a successful paradigm to understand fracture behavior by rationalizing the phenomenology of material failure in presence of intentional flaws or inherent defects at continuum level. It has provided a most effective description of extremely brittle solids such as silica glasses and ceramics. Besides brittle steels, linear elastic fracture mechanics (LEFM) has also been applied[1-3] to characterize fracture behavior of brittle glassy polymers.[4-6] On one hand, these textbooks[4,5] assume that observed brittle fracture[7,8] in uncut polymers takes place prematurely at breaking stress $\sigma_b$, caused by stress intensification arising from inherent flaws, in absence of which brittle polymers would have shown significantly higher inherent fracture strength $\sigma_{F(inh)}$ ($>> \sigma_b$). On the other hand, fracture mechanics has inspired molecular design to achieve rubber toughening[9] of polystyrene (PS) and polymethyl methacrylate (PMMA).

Before a literature survey on fracture behavior of PS and PMMA, we first indicate effects that are beyond fracture mechanics description. For example, we have to leave it to polymer physics to answer why physical aging turns a ductile glassy polymer brittle,[10,11] why hydrostatic pressure does the opposite,[12] why brittle polymers no longer undergo brittle fracture after melt stretching[13,14] or why pre-melt-stretched ductile polymer appears brittle when drawn perpendicular to the melt-stretching direction,[11] why mechanical rejuvenation makes a brittle polymer ductile.[15] There are other deep questions such as why crazing[16,17] arises,[18] why glassy polymers can yield in presence of crazing as well as when, how and why they undergo brittle-ductile transition (BDT) over a narrow temperature window. Until recently,[14,18] the Ludwig-Davidenkov-Orowan (LDO) hypothesis[4,6] has been regarded as a standard way to rationalize BDT; on the other hand, lack of ductility and appearance of crazing has been said to arise from insufficient entanglement.[16,19] The statement[6] that "polymers are intrinsically brittle solids and fracture in a brittle manner at low temperatures and/or high strain rates" is actually a paraphrase of the LDO hypothesis.

According to the recent theoretical considerations based on a coherent analysis of available phenomenology,[14] chain networking due to interchain uncrossability is the driver for molecular activation and ductility while no predictive description exists[18] of short-ranged intersegmental interactions. Bisphenol A polycarbonate (PC) is ductile, whereas polystyrene (PS) is brittle at room temperature not because PC is more "entangled" than PS, but because PC apparently can more readily undergo activation in its glassy state. At their BDT, all glassy polymers made of linear flexible chains, such as PS and PC, presumably have the same capacity to bring about activation.[18] According to this phenomenological model,[14] inherent fracture strength $\sigma_{F(inh)}$ at BDT scales linearly with the areal density $\psi_{LBS}$ of load-bearing strands (LBS) that characterizes the structure of chain networking. Consequently the classic Vincent plot[7] acquired a new interpretation: $\sigma_{F(inh)} = \psi_{LBS}f_{cp}$ at BDT due to chain pullout at critical force $f_{cp}$ rather than chain scission.[20] In this plot, thirteen different polymers show the breaking stress $\sigma_b$ (fracture strength) to scale linearly with the areal density $\phi$ of backbone bonds, i.e. $\sigma_b \sim \phi$. Because $\phi$ can be shown[14] to be proportional to $\psi_{LBS}$, the Vincent plot hints that $\sigma_b \sim \sigma_{F(inh)}$ if $f_{cp}$ is the same for the different polymers. Thus, we have clarified such statements as[6] "it must be concluded that during the fracture of these 13 polymers, on average, much less than 1% of the fracture area involves C−C bond scission". According to Ref. [14], flaw-free uncut specimens would show brittle fracture because the chain network is unable to retain its structural integrity during its attempt to bring about activation below BDT.

---

[*] Corresponding author at swang@uakron.edu

The conventional wisdom in the literature[5] states that "…the fracture strength ($\sigma_b$) will be controlled principally by the size of the largest cracks or flaws in the structure. Hence, if the flaw dimensions are known, $\sigma_b$ can be predicted" from $G_{Ic}$; "Several brittle polymers behave as if they contain flaws of a particular size…"; "…it is obvious that increasing the amount of plastic deformation at the crack tip will have the required effect" of toughening polymers. It is standard to regard the critical energy release rate $G_{Ic}$ as a material constant capable of prescribing fracture strength of uncut specimens. To reiterate, it is widely known consensus that "the level of strength of many materials is well below the theoretical strength, or inherent strength, $\sigma_{F(inh)}$, due to the presence of defects", "the strength of the polymer increases as the crack length is reduced",[21] and "the real strength of the material is much less because defects exist within the material that result in very highly stressed local areas."[8] Based on Berry's observations that through-cut of size smaller than $a^*$ shows comparable breaking stress to $\sigma_b$ observed of uncut PS and PMMA, Haward[21] concluded that such polymers show lower fracture strength $\sigma_b$ because "they behave as though they contain natural flaws of these critical sizes ($a^*$) that are termed 'inherent' or 'intrinsic' flaws".

Fracture mechanics characterizes fracture behavior based on the global stress state and identifies $G_{Ic}$ as an essential property to rank materials. To verify this correlation between local and global states, there have been numerous photoelasticity studies in the literature[22-24] to show that the stress intensity factor K can indeed be prescribed by the LEFM analysis of Westergaard and Irwin[25, 26]. For example, isochromatic fringe orders have been used[22, 23, 27-29] to obtain K. The method is particularly convenient for the study of dynamic fracture.[30] For PMMA, the photoelastic patterns allowed Green and Pratt[30] to reveal a range of 1 to 10 MPa.m$^{1/2}$ for $K_c$, corresponding to different crack propagation velocities. Using polysulfone, Gales and Mills[31] showed that the photoelastic measurement of K agrees well with the theoretical formula for $K_I \sim (\pi a)^{1/2}\sigma$ for mode I (tensile opening). However, these past photoelasticity studies[22] as well as other investigations[32-34] typically did not quantify how the local stress changes as a function of distance from the cup tip and how the stress level at the crack tip correlate with applied load upon fracture.

In this publication we apply birefringence measurements to examine Haward's conclusion by using fracture mechanical analysis as a tool to characterize fracture behavior of glassy polymers. Specifically, we gather evidence to compare with the prevailing view that mechanical strength of brittle polymers is determined by characteristics of inherent flaws. The present paper 1 focuses on failure behavior of glassy polymers and shows that toughness (critical stress intensity factor $K_{Ic}$) is determined by $\sigma_{F(inh)}$ and a material-specific length scale P. In our subsequent publications[35, 36] we will determine whether a similar description applies in the description of brittle fracture behavior of elastomers.

## 2. Theoretical considerations and analyses

In this section, after a brief mention of BDT we first present the conventional fracture mechanics description relating the critical load $\sigma_c$ to the crack size $a$. We then discuss the difficulty associated with it as it relates to fracture behavior of plastics. Finally, we propose a plausible viewpoint and experimental tests to verify this scenario.

*2.1 Nature of BDT*

In materials such as plastics, especially amorphous glassy polymers, the existence of a rather distinct BDT requires investigations beyond application of fracture mechanics. Here mere application of Eyring's idea of stress-induced activation[37] is insufficient: According to this argument of stress-induced activation, a polymer would always stay ductile independent of temperature because the activation barrier would always be lowered by imposing sufficiently high stress. According to a recent phenomenology-based molecular model for yielding and BDT in glassy polymers,[14] a glassy polymer can yield only when the chain network is robust enough to transmit the stress (molecular force to be more precise) required for activation. Otherwise, e.g., below BDT it shows brittle fracture because the network due to chain uncrossability breaks down before producing global activation. In other words, we must explain what structure supports and transmits mechanical stress before applying Eyring's activation concept.

*2.2 Nominal (fracture) strength vs. inherent strength: are there intrinsic flaws?*

In the current literature, whenever brittle fracture is observed in uncut specimens, the nominal breaking stress (fracture strength) $\sigma_b$ is uniformly assigned a fracture mechanical origin: Fracture arises from sizable "intrinsic flaws" so that $\sigma_b$ may be much lower than inherent fracture strength $\sigma_{F(inh)}$. It requires three assumptions for $\sigma_{F(inh)} >> \sigma_b$ to hold: (a) There are intrinsic flaws of considerable size $a^*$; (b) the effect of such flaws is the same as that of through-cracks of size $a^*$; (c) fracture is preceded by emergence of a yielding or plastic zone of size $r_p^* << a^*$, i.e., stress buildup at crack tip continues until $r_p^*$.

It is standard practice to identify a threshold $a^*$ below which any further decrease of cut length $a$ no longer increases the critical load $\sigma_c$ at fracture of such a precut specimen. In other words, one can use a precut specimen with $a \leq a^*$ to show that $\sigma_c$ becomes comparable to $\sigma_b$. In the logic of fracture mechanics, there must be intrinsic flaws to cause such an uncut specimen to exhibit the same strength as that of the precut specimen with cut length $a^*$. For simplicity, in this paper, our analysis will neglect small deviation from the idealized limit of infinite elastic body and consider only mode I (tensile opening mode) loading. In terms of $a^*$ and nominal load $\sigma_0$ (i.e., far-field stress) the stress intensity factor[4, 5, 26, 38]

$$K_I = \sigma_0(\pi a)^{1/2} \tag{1}$$

is related at fracture or failure to other quantities in two different forms:

$$K_{Ic} = \sigma_b(\pi a^*)^{1/2} = \sigma_{F(inh)}(2\pi r_p^*)^{1/2} \text{ for brittle fracture} \tag{2a}$$

and

$K_{Iy} = \sigma_y(\pi a^*)^{1/2} = \sigma_{Y(inh)}(2\pi r_p^*)^{1/2}$ for ductile failure
(yielding/necking), (2b)

where $\sigma_{Y(inh)}$ denotes inherent yield strength: for ductile polymers, a small enough cut at $a^*$ should permit us to observe the same yield stress $\sigma_y$ as measured from uncut samples.

It is necessary to emphasize that while the first equalities in Eqs. (2a) and (2b) capture the meaning of $a^*$ in conventional fracture mechanical account, the second equalities are our extrapolation regarding a crucial connection that the stress intensity approach is supposed to make by assigning the proper meaning of $K_{Ic}$ or $K_{Iy}$ (critical $K_I$ corresponding to onset of yielding in ductile fracture). In the literature of fracture mechanics, only the second equality in Eq. (2b) is familiar, usually presented as an account for brittle fracture. Therefore, the second equality in Eq. (2a) may be regarded as our revision to encompass the possibility of absence of yielding in brittle materials. We note that $\sigma_{F(inh)} \gg \sigma_b$ or $\sigma_{Y(inh)} \gg \sigma_y$ could take place only if $r_p^* \ll a^*$. Conversely, $\sigma_b$ or $\sigma_y$ would be comparable to $\sigma_{F(inh)}$ or $\sigma_{Y(inh)}$ if $a^*$ and $r_p^*$ are comparable. Since $a$ is always much smaller than the specimen width W, the finite numerical corrections[38] of the order of $a$/W is ignored in our analysis.

At least for glassy polymers such as PS and PMMA that are optically clear, the suggestion of inherent flaws of size $a^*$ is not supported by visual inspection. Berry reported[2] $a^* \sim 1$ mm for PS because PS with precut size $a < 1$ mm was found to be as strong as uncut PS. The identification of Berry cut length $a^*$ as revealing flaws of such size has been taken to advance the argument like "such flaws, however, are not present in the material before deformation but appear to be formed on loading"[21] and "these flaws are not present in the material before deformation but appear to be formed during loading."[5] The flaws, referred in these textbooks, are the so-called crazes. Thus, crazes have long been regarded as the precursor to brittle fracture. Here we meet a difficulty in logic: Flaws are only invisibly small in undeformed polymers, and crazes only form during drawing and cannot act like through-cuts. Crazes also occur above BDT and are thus not a sufficient condition for brittle fracture. They are the natural structural responses of various glassy polymers to large tensile extension, as explained below. Irwin's concept of plastic zone[39] may not apply to brittle glassy polymers. For polymers that are transparent, the manifestation of yielding zone is perhaps amenable to birefringence measurements.

### 2.3 Crazing is not inherent flaw

Craze[16] formation involves an activated process in the sense that some glassy polymers take longer time to undergo the localized decohesion before brittle fracture via catastrophic breakdown of the chain network.[18] The loading time and the level of loading determine the crazing characteristics such as craze size and density at a given temperature. In other words, the crazing intensity depends on the external condition and on where a glassy polymer settles on its energy landscape. For example, crazing would not occur below a certain strain or stress; under a given load for a specified period, freshly prepared amorphous poly(lactic acid) (PLA) is craze-free at room temperature while the same sample under the same external condition shows crazing after physical aging. There is obviously no addition or abstraction of flaws in annealing. However, physical aging has made it more difficult for molecular activation to take place in aged PLA. In absence of sufficient molecular mobility, thanks to physical aging, PLA is unstable against volume expansion arising from its inability to undergo contraction transverse to the imposed tensile extension. Cavitation in the form of crazing emerges at locations of molecular "defects" or "heterogeneities" where the "horserace produces winners": The chain network presumably locally collapses through chain pullout involving LBSs with high chain tension, creating strain localization in the sense that much higher tensile strain occurs at various locations. Tensile strain necessarily requires transverse contraction. However, an overall shape change by global transverse contraction cannot occur without sufficient global molecular mobility. Without molecular activation to permit contraction, cavitation appears, with fibril bundles of polymer chains occupying a fraction of the volume in a given cavity. Such a configuration on the scales of microns or higher, scatters light and is known as crazing. In this sense, the appearance of crazes does not really require the system to be inherently heterogeneous on length scales that characterize crazing.

It is helpful to summarize several points regarding crazes in uncut glassy polymers: (a) Before physical aging, a glassy polymer such as PLA is ductile and free of crazing. In the simplest scenario, physical aging does one thing, i.e., bringing the PLA to a lower energy state on the energy landscape to make it more difficult for deformation-induced activation to take place. The aged PLA is brittle and prone to crazing. (b) Aged PET also offers us a chance to examine the influence of crazing on fracture – PET can draw considerably in presence of massive crazing. (c) Crazing and brittle fracture have everything to do with the failure to activate the glassy state. This is the leading-order physics. The question of what enables activation and yielding in glassy polymers has been addressed.[14, 18] (d) Adequate pre-melt-stretching eliminates crazing and brittle fracture, supporting the theoretical picture concerning why yielding can take place in ductile polymers: In the recently proposed model,[14] chain network has been recognized as the driver to cause yielding by bringing about activation. Pre-melt stretching makes it possible for otherwise brittle polymers to attain activation and minimize the chance of craze formation.

### 2.4 Stress intensification analysis

Stress buildup around a through-crack in an infinite linear-elastic body was first discussed by Westergaard[25] and subsequently by Irwin[26]. The stress components are found to have the following forms for tensile extension along y axis,[38]

$\sigma_{yy}(r) = K_I f(\theta)/(2\pi r)^{1/2} + \sigma_0$, $\sigma_{xx}(r) = K_I g(\theta)/(2\pi r)^{1/2}$,
$\sigma_{xy} = K_I h(\theta)/(2\pi r)^{1/2}$, $\sigma_{zz} = 0$ (plane stress) or $\nu(\sigma_{xx} + \sigma_{yy})$ (plane strain), $\sigma_{zx} = \sigma_{yz} = 0$, (3a) where $f(\theta)$, $g$ and $h$ are explicit functions of the angle formed by the position vector **r** (originating from the crack tip) with the x-axis (direction of crack propagation). Here we have added the far-field stress $\sigma_0$ so that $\sigma_{yy} - \sigma_{xx}$ approaches $\sigma_0$ for $r \gg a$. For subsequent discussion in Section 2.7, we write the explicit expressions for the functions

$f(\theta) = \cos(\theta/2)[1+\sin(\theta/2)\sin(3\theta/2)]$,
$h(\theta) = \cos(\theta/2)\sin(\theta/2)\cos(3\theta/2)$,
and $(f - g) = 2\cos(\theta/2)\sin(\theta/2)\sin(3\theta/2)$. (3b)

Eq. (3a) suggests that fracture would occur at arbitrarily small $K_I$ because the stress at the crack tip ($r = 0$) can be arbitrarily high. Since fracture obviously does not occur at vanishing $K_I$, Kinloch concluded[5] "stress alone does not make a reasonable local fracture criterion". Such understanding questioned the essence of the stress intensity approach. To deal with the stress divergence in Eq. (3a), a failure (plastic) zone is envisioned in standard fracture mechanics treatment. In such a zone the stress ceases to increase because the material cannot withstand higher stress than some inherent yield strength $\sigma_{Y(inh)}$. According to fracture mechanics, for a given through-cut length $a$ at an applied tensile stress $\sigma$, the zone size $r_p$ grows[5, 38] with the load parameter $K_I$ of Eq. (1) as[5, 38]

$$r_p = (1/2\pi)(K_I/\sigma_{Y(inh)})^2. \qquad (4)$$

Since $\sigma_{Y(inh)}$ is a material constant, $r_p$ simply increases in proportion to $K_I^2$.

According to textbook, fracture occurs at $K_{Ic}$ when the zone size has grown to a critical value $r_p^*$. However, there is no instruction from fracture mechanics as to why $r_p$ must increase with $K_I$ to $r_p^*$ for fracture to take place, given by

$$r_p^* = (1/2\pi)(K_{Ic}/\sigma_{Y(inh)})^2. \qquad (5)$$

In other words, the introduction of the concept of plastic zone has not allowed the local stress analysis (as one pillar in fracture mechanics) to describe why fracture occurs. Eq. (5) is Eq. (2a). However, it does not prescribe $K_{Ic}$ because there is no separate argument to estimate $r_p^*$. Consequently, the energy balance argument of Griffith[40] takes center stage in fracture mechanics. In passing, we note that it is rather unusual to suggest that brittle fracture is preceded by formation of a sizable plastic zone for glassy polymers well below BDT where they are incapable of undergoing plastic deformation. In other words, since PS and PMMA are brittle at room temperature presumably because of the lack of molecular activation, it is unlikely that yielding and plastic deformation emerge before brittle fracture.

*2.5 Energy balance argument*

According to the energy balance argument (the other pillar) in fracture mechanics, a linear-elastic material needs to store enough energy from mechanical deformation to pay for the energy required to create new surface associated with crack propagation. Under mode I loading, the energy gain per unit surface at a given nominal load in presence of a through-crack of size $a$, known as the energy release rate $G_I$ can be shown to be of a form expressible in terms of $K_I$ of Eq. (1) as $K_I^2/E$, with E being the Young's modulus. Griffith[40] originally proposed for silica glasses that fracture occurs when $G_I$ approaches a critical value $G_{Ic}$ greater than surface fracture energy $\Gamma_f$. Here $G_{Ic}$ can be measured in terms of $\sigma_c$

$$G_{Ic} = K_{Ic}^2/E = \pi\sigma_c^2 a/E = w^c(2\pi a), \qquad (6a)$$

where $w^c$ as a work density is given by

$$w^c = \sigma_c^2/2E. \qquad (6b)$$

We note that the form given by the third equality in Eq. (6a) resemble the Rivlin-Thomas[41] expression for $G_{Ic}$ for fracture in pure shear of elastomers.

Alternatively, for a crack to become unstable, i.e., for it to advance, $K_I$ goes to

$$K_{Ic} = \sigma_c(\pi a)^{1/2} \qquad (7)$$

where $\sigma_c$ can be computed from Eq. (6a-b) with $G_{Ic} = \Gamma_f$ for Griffith case of glasses. Since ever $G_{Ic}$ was found to be larger than $\Gamma$ for other materials, Irwin[42] suggested to treat $G_{Ic}$ as a quantity that encompasses plastic dissipative processes so that we can still express fracture strength $\sigma_b$ for other materials using the Griffith criterion for silica glasses:

$$\sigma_c = (EG_{Ic}/\pi a)^{1/2}, \qquad (8)$$

i.e., $G_{Ic}$ is recognized as the key parameter to estimate in theory[17, 43] and to measure in experiment[1, 2]. In fact, it is perhaps more instructive to figure out why $K_{Ic}$ of Eq. (7) is constant at fracture.

For fracture of brittle inorganic glasses, the critical energy release rate $G_c$ reveals the fracture surface energy $\Gamma_f$. However, using through-cut, Berry[2] found the brittle PS and PMMA to have $G_{Ic} \sim 1.7$ kJ/m$^2$ and 0.3 kJ/m$^2$ respectively, orders of magnitude greater than $\Gamma_f \sim$ ca. 1 J/m$^2$, which can be estimated by assuming cleavage of all covalent bonds across the fracture surface: Given the molecular cross-section of a polymer bond,[44] $s = pl_K$, where $p$ and $l_K$ are the packing and Kuhn length respectively whose values are well known for PS and PMMA,[45] and $E_b = 300$ kJ/mol as the dissociation energy, we have $\Gamma_f \sim E_b/s = 0.8$ J/m$^2$, taking $p = 4$ Å and $l_K = 1$ nm. It has been presented elsewhere that for glassy polymers brittle fracture plausibly involve chain pullout in a chain network instead of scission.[20] Thus, such a textbook estimate[46] may be a huge overestimate. In face of the irreconcilable discrepancy of three orders of magnitude, the standard explanation is that fracture is largely dominated by plastic dissipative processes. However, we still face the challenge to explain why $G_{Ic}$ is on the order of 1 kJ/m$^2$ for these polymers.

Since the four stresses in Eqs. (2a-b) are material properties of uncut samples, we can deduce that $a^*$ and $r_p^*$ are proportional to each other. Before closing on the review of fracture mechanics of plastics, it is necessary to mention a third length scale known[47, 48] in fracture mechanics as the fractocohesive length $L_{fc}$, given by the ratio of $G_{Ic}$ to the work of fracture $w_c$

$$L_{fc} = G_{Ic}/w_c, \qquad (9a)$$

where $w_c$ is measured using an uncut specimen in terms of $\sigma_b$ as

$$w_c = (\sigma_b)^2/2E. \qquad (9b)$$

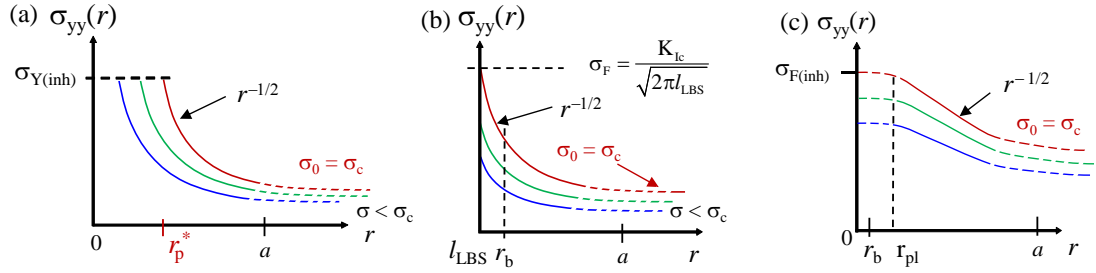

**Fig. 1.** (a) Standard viewpoint of stress intensification at a through-crack tip at several loads including the critical load $\sigma_c$ at fracture, with tip stress saturating to inherent yield strength at $r_p$, which characterizes the size of the plastic zone and grows with $K_I$ according to Eq. (4). (b) A revised scenario where the tip stress builds to an artificial value $\sigma_F$ all the way to the mesh size of the chain network, $l_{LBS}$, which can be expected to be only a few nanometers. (c) A case contrasting (a) and (b), which can be verified when the zone of stress plateau has a size $r_{pl}$ greater than the spatial resolution $r_b$.

In Section 5 we will discuss the relationship among the three lengths $a^*$, $r_p^*$ and $L_{fc}$ within the fracture mechanics paradigm.

*2.6 More realistic scenario for polymers*

According to the textbooks, $r_p$ would grow with $K_I$, as shown in Eq. (4), while the stress level in the yielding zone would remain saturated at $\sigma_{Y(inh)}$, as described in the preceding Section 2.4 and explicitly summarized in Fig. 1a. Here it is important to emphasize that both $r_p^*$ and $\sigma_{Y(inh)}$ are unknown within the analysis. Similarly, Eq. (4) does not even prescribe the magnitude of $r_p$ for a given value of $K_I$ since $\sigma_{Y(inh)}$ is not unknown. On the other hand, if $\sigma_y \ll \sigma_{Y(inh)}$, it would imply that $r_p^* \ll a^*$. Moreover, for an amorphous polymer below BDT, it remains to be shown how a plastic zone would emerge before brittle fracture. This difficulty can be traced back to the need to address the issue of stress divergence in Eq. (3a).

Alternatively, for glassy polymers, the characteristic size $l_{LBS}$ of load-bearing strands of the chain network introduces a natural cutoff. Thus, we do not expect LEFM to apply below the length scale of $l_{LBS}$. In other words, the plastic zone cannot have a size smaller than $l_{LBS}$. If the stress intensification would occur all the way to $l_{LBS}$, Eq. (3a) would anticipate the local stress to reach a level higher than the load $\sigma_0$ by a factor of $(a/2l_{LBS})^{1/2}$. In other words, in place of Fig. 1a, we would have a modified scenario like Fig. 1b. In doing so, we eliminate one unknown parameter, i.e., $r_p$, along with the need to quantify the fracture condition in terms of $\sigma_{Y(inh)}$ of brittle materials. Instead, the stress buildup stops at the cut tip to a level given by the horizontal dashed line at $\sigma_{F(inh)}$. Unlike Fig. 1a, the fracture criterion is unambiguous: The critical load is given by $\sigma_c = \sigma_{F(inh)}(l_{LBS}/a)^{1/2}$. Given Berry's observation[1, 2] of brittle fracture in PMMA at $\sigma_c = 20$ MPa for $a = 1$ mm, the scenario depicted in Fig. 1b suggests $\sigma_{F(inh)} = 6.3$ GPa, given $l_{LBS} =$ ca. 5 nm. This magnitude of 6.3 GPa is higher than $\sigma_b$ reported by Vincent[7] at the BDT by a factor of 100. As indicated in Section I, $\sigma_{F(inh)}$ may be only as high as $\sigma_b$: The factor of 100 stems from the fact the chain pullout force may be only one tenth of the breaking force of a covalent bond and the areal density $\psi_{LBS}$ of load-bearing strands may be only one tenth of the bond areal density $\phi$.

Could the stress intensification die off on the scale of $r_b$? In other words, at least for plastics, can Fig. 1c be a more realistic depiction of the stress field for a pre-cut specimen? If stress ceases to grow at $r_{pl} > r_b$, we will be able to determine the tip stress at fracture, which can be taken to represent the inherent fracture strength $\sigma_{F(inh)}$. We can in fact quantitatively determine the relations given by Eqs. (2a-b), i.e., rewriting them as

$$\sigma_{F(inh)} = \sigma_b S \text{ and } \sigma_{Y(inh)} = \sigma_y S, \ S = (a^*/2r_{pl})^n \qquad (10)$$

where a more generalized form is adopted, with $n = ½$ corresponding to the Westergaard[25]-Irwin[26] result. Here we may call the dimensionless $S$ the stress intensification. In this scenario, we will be compelled to ask why stress ceases to grow beyond $r_{pl}$ and what determines the magnitude of $r_{pl}$ for a given polymer. Since the stress level-off is depicted to occur below the critical load $\sigma_c$ for fracture and the tip stress increases progressively with $\sigma_0$, this stress plateau region has nothing in common with the concept of plastic zone depicted in Fig. 1a.

While the picture of Fig. 1a does not assert that $\sigma_{Y(inh)} \gg \sigma_b$, the modified scheme in Fig. 1b encompasses the prevailing view that $\sigma_{F(inh)} \gg \sigma_b$ because $\sigma_{F(inh)} = \sigma_b(a^*/2l_{LBS})^{1/2}$, where use is made of Eq. (2a). Taking, $a^* \sim 1$ mm for PS and $l_{LBS} = 5$ nm, we have $\sigma_{F(inh)} = 300\sigma_b$. However, our estimate[14] shows that this is unlikely the case. On the other hand, to have $\sigma_b \sim \sigma_{F(inh)}$ below BDT and $\sigma_y \sim \sigma_{Y(inh)}$ above BDT, we would infer $a^* \sim 2r_{pl}$ in Eq. (10), i.e., the two length scales sharing the same origin. Fig. 1c encompasses this possibility. Validation of Fig. 1c does depend on having sufficient spatial resolution, i.e., having $r_b < r_{pl}$. By Fig. 1c, we propose that brittle fracture takes place because the tip stress has reached $\sigma_{F(inh)}$. Rewriting Eq. (3a), we have

$$\sigma_{F(inh)} \sim K_{Ic}/(2\pi r_{pl})^{1/2} \qquad (11)$$

which implies that the elusive magnitude of $K_{Ic}$ is actually determined by $\sigma_{F(inh)}$ and $r_{pl}$.

*2.7 Experimental estimate of inherent strength $\sigma_{F(inh)}$ or $\sigma_{Y(inh)}$ from birefringence*

Our recent model hints that inherent fracture or yield strength $\sigma_{F(inh)}$ or $\sigma_{Y(inh)}$ may not be much higher than $\sigma_b$ or $\sigma_y$. Since there exists a stress/strain optical rule (SOR) relating the applied strain and corresponding birefringence for glassy polymers,[31, 34, 49, 50] as illustrated by the solid line in Fig. 2, the setup shown in Fig. 3 may be used to reveal the retardance near a cut tip. Within the spatial resolution ($r_b$) of birefringence observations, we can determine the local stress level. Stress plateau would be observed if $r_b < r_{pl}$. These birefringence experiments can be carried out either below or above BDT. Alternatively, we can use both brittle and ductile polymers at room temperature to learn about how fracture behavior differs, e.g., between PMMA and PET.

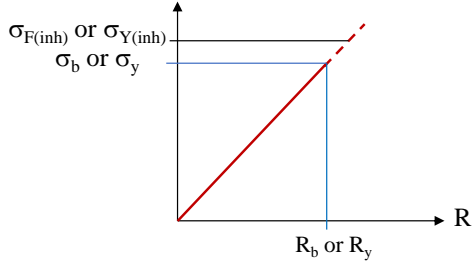

**Fig. 2.** Illustration of stress-optical rule, correlating the retardance R with the tensile stress, which may be obtained using the birefringence setup shown in Fig. 3 up to brittle fracture or yielding at $R_b$ and $R_y$ or $\sigma_b$ and $\sigma_y$ respectively.

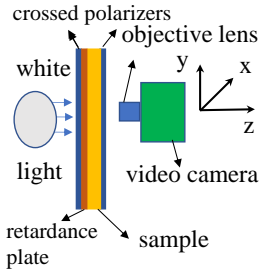

**Fig. 3.** Sketch of a birefringence setup based on white light for measurement of spatial retardance field, involving two crossed polarizers and a retardation plate that is either arranged to cancel or add to the emergent retardance due to drawing along y axis.

Mechanical deformation of glassy polymers produces segmental orientation, leading to optical birefringence. For example, for uniaxial extension, the principal stresses are along and perpendicular to the direction of drawing (y-axis) so that a stress-optical rule (SOR) to relate the birefringence $\Delta n$ is proportional to the tensile stress T

$$\Delta n = CT \tag{12}$$

where C is the stress-optical constant. Tensile extension of precut stripes involves either plane stress or plane strain at cut tip. It is straightforward to diagonalize the stress tensor, i.e., to identify the principal stresses $\sigma_1$ and $\sigma_2$.[51] Birefringence measurements are ideally suitable to quantify $\Delta\sigma = \sigma_1 - \sigma_2$ around cut tip: $\Delta n(r) = C\Delta\sigma$. A non-zero shear stress $\sigma_{xy}$ rotates the principal stress direction away from the tensile direction (y-axis) in contrast to uniaxial extension, to an angle given by $\tan\chi = 2\sigma_{xy}/(\sigma_{yy} - \sigma_{xx})$. The principal stress difference is given by

$$\Delta\sigma = [(\sigma_{yy} - \sigma_{xx})^2 + 4(\sigma_{xy})^2]^{1/2}. \tag{13}$$

Inserting the expressions in Eq. (3a) and Eq. (3b) into Eq. (13) and solve for the tensile stress $T = \sigma_{yy} - \sigma_{xx}$, we have

$$T = \sigma_{yy} - \sigma_{xx} = \{[(1 + A^2)\Delta\sigma - A^2\sigma_0^2]^{1/2} + A^2\sigma_0\}/(1 + A^2) \tag{14a}$$

where $A(\theta)$ is given by

$$A(\theta) = 2h/(f - g) = \cot(3\theta/2) \tag{14b}$$

At $\theta = \pi/3$, $A(\theta) = 0$ so that Eq. (14a) simplifies to yield

$$T = \Delta\sigma = \Delta n/C \tag{14c}$$

which follows from Eq. (12). In combination with Eqs. (3a-b), we can rewrite Eq. (14c) as

$$K_I/(2\pi r)^{1/2} = (2/3^{1/2})(\Delta n - \Delta n_0)/C. \tag{14d}$$

In other words, Eq. (14d) shows the birefringence measurements can be carried out to obtained $K_I$, as done before.[31]

*2.8 Circular cracks*

When $r_{pl} < r_b$ it would be challenging to discern any difference between Fig. 1b and Fig. 1c. In this case, it may still be possible to learn about failure behavior through characterization of local stress field by considering a different crack type. For example a circular through-crack (hole of radius *a*) shows[52, 53] stress intensification on a length scale proportional to *a*. Thus, unlike single-edge notch (SEN), by having a circular hole of large enough size, e.g., $0.1a > r_b$, we can capture the stress level at failure without the resolution constraint. According the LEFM analysis, the stress buildup goes as

$$\sigma_{zz}(r, \theta) = (\sigma_0/2)\{1 + (a/r)^2 - [1+3(a/r)^4]\cos 2\theta\}^2 \tag{15}$$

so that the stress at $r = 1.1a$ is only lower than that at $r = 0$ by a known amount of 19% at the two poles. Here poles locate at $\theta = \pi/2$, where $\theta$ is the angle that the position vector **r** makes with the drawing direction (z-axis), and $\sigma$ is the far-field stress. Upon resolving the retardation at $r = 1.1a$, i.e., $0.1a$ from the edge of the cut and reading the birefringence at the two poles at the moment of failure, we should be able to determine the local stress level. In other words, with a circular through-cut, the stress intensification occurs on a prescribed length scale that can be chosen to be sufficiently large instead of an unknown scale of $r_{pl}$ that could be below $r_b$.

### 3. Methods and Materials
*3.1. Sample Preparation*

Three polymers were studied in this work. Table 1 lists the basic information about these samples, which were supplied as sheets.

Dogbone and stripe-shaped specimens were prepared by first tracing a dogbone design onto the sheets and then either, for PET and PC, cutting with scissors and paper trimmer to carefully to avoid introducing substantial edge defects or, for PMMA, removing excess material with a coarse sanding belt and then smoothing the edges with a flat file. Dogbone-shaped specimens were employed to obtain the stress-optical relationship for PET and PC and stripe-shaped specimens to study the effect of single-edge notch and circular holes.

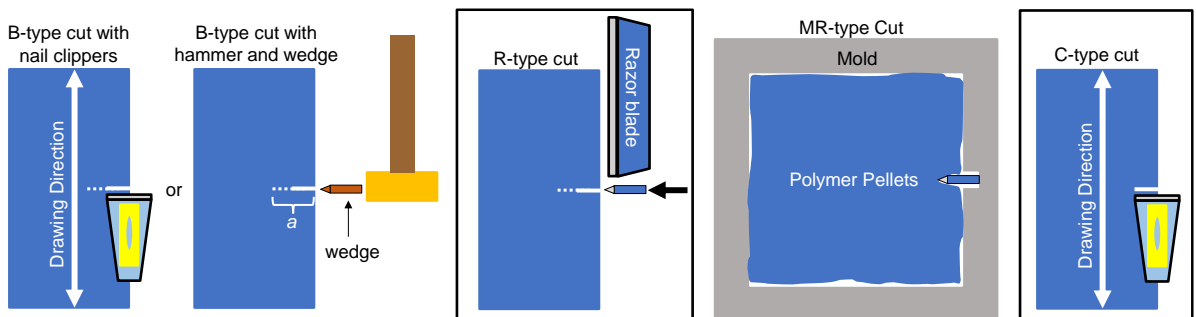

**Table 1.** Specimen Characteristics

| Samples | Sheet Thicknesses (mm) | Source |
|---|---|---|
| PET | 0.25 | Auriga Polymers Inc. |
| PC | 0.25 | Grainger LLC |
| PMMA | 3.0, 5.5 | Professional Plastics |

**Fig. 4.** Single-edge notch (SEN) was introduced to specimens by four methods. (1) B-type (similar to that of Berry[1]): Rapid application of force with nail clippers or gentle hammering of razor blade or wedge against the specimen edge generated a thin crack that spontaneously propagated further across the sheet. If the crack did not propagate more-or-less parallel to the width direction, the specimen was discarded. (2) R-type: Hammering a razor blade into the sheet's edge to generate a notch tip that was defined by the tip of the razor blade. (3) MR-type: Melt compression of polymer pellets within a squared-shaped steel mold that had a razor blade embedded into a slot along the side of the mold produced an R-type notch that did not suffer from residual birefringence caused by notch preparation. (4) C-type: pre-heated specimen was cut with pre-heated nail clippers, then thermally annealed to reduce residual birefringence near the notch tip.

SEN was introduced by one of four methods, as described in Fig. 4. Most specimen notches were B- and R-type. After cutting, specimens were inspected under magnification between cross polarizers. Specimens with overt defects along the notch tip or circular hole were discarded. The remainder were either directly tested despite some residual birefringence near the crack tip or were annealed slightly above $T_g$ either in an oven or with an industrial blow dryer to partially heal the region near the notch tip in order to remove the residual birefringence produced by cutting. When necessary, specimens were held under constraint to minimize warping during annealing.

Specimens with B-type cracks were cut while below the BDT. Crack was introduced in PMMA specimens at room temperature by hammering a glass-scrapper against the side of the sheet. PET specimens were chilled to below room temperature either in a freezer (-20 °C) for 15-30 minutes or liquid nitrogen (LN$_2$) for a minute, then cut with a similarly chilled nail clipper while still in the freezer or submerged in LN$_2$. PC specimens were chilled in LN$_2$ for a minute and then cut with a nail clipper while submerged.

PET specimens with R-type notch were cut at room temperature and annealed at 75-90 °C for 30-60 minutes to remove the residual birefringence incurred during cutting. The degree of residual birefringence in PET immediately after cutting was on the order of R = 500-600 nm; thermal annealing reduced this to ~100 nm over the course of 30-60 minutes at annealing temperatures within a few degrees of Tg. Further time annealing did not further reduce the degree of residual birefringence.

PMMA with R-type cuts were prepared by sawing a notch along the specimen edge with either a razor blade or bandsaw blade, and were annealed as needed to remove the residual birefringence incurred during cutting.

PMMA specimens with MR-type notch were prepared through press-molding of pellets (ALTUGLAS®, supplied by Arkema) at 180 °C under a load of 15 000 lbs for 20 minutes before allowing the specimens to cool while under compression for 10 minutes and subsequently removing them to cool to room temperature over a period of several minutes. Press-molding was performed in a mold that had a thin slit which would accommodate a razor blade, as shown in Fig. 4, and that would result in a notch whose sharpness is dictated by the razor blade employed.

Lastly, PMMA specimens with C-type cut were prepared by annealing uncut press-molded PMMA sheets at 130 °C for 15 minutes, cutting with a pre-heated nail clipper, and then annealing for another 15 minutes to heal the region in front of the crack tip.

Circular cracks were introduced to specimens at room temperature using a drill press. Specimens were affixed with clamps and a single layer of sacrificial 0.75 mm PC films were placed between the specimen and clamps to avoid localized damage to the specimen from clamping. A standard black oxide drill bit rotating at 620 rpm was gently lowered onto the taut specimen until penetration. Excess material along the edge of the circular crack was gently removed with a razor blade to avoid further deformation near the crack edge.

*3.2. Methods*

Tensile extension of uncut and cut specimens was carried out at room temperature on an Instron 5969 tensile tester between crossed polarizer films (obtained from Polarization.com). The setup is illustrated in Fig. 3. The reported draw ratio $L/L_0$ is based on the initial length $L_0$ of the narrow section of the dogbone specimens and the inter-clamp distance for the stripe specimens.

The setup of birefringence measurement in Fig. 3 allows us to observe the evolution of colors due to increasing birefringence according to a Michel-Lévy chart, such as that reproduced in Fig. 5a from an online source. Here the color-rich second order is ideally suited to observe small changes in birefringence. A retardation plate (prepared with the machine direction (MD) perpendicular to the drawing direction – y-axis) from a group of 0.75 mm thick PC sheets (obtained from Grainger LLC) is negative, each producing an amount of shift in retardation on the order of −300 nm. Different combinations of these PC sheets were used to have various starting retardance as illustrated in Fig. 5a by the vertical lines. Specifically, since PMMA shows negative birefringence at room temperature,[54, 55] placing such a negative retardation plate allows the birefringence setup (cf. Fig. 3) to reveal retardation traveling rightward in the ML chart during drawing, as marked in Fig. 5a and illustrated in Fig. 5c. Given the positive birefringence of PET and PC specimens, the use of negative retardation plate causes the retardation to travel during drawing, as indicated in Fig. 5a and illustrated in Figs. 5d and 5e.

In-situ videos were recorded with either an iPhone camera connected to a clip-on macro-lens to facilitate inspection of specimen at close-range or a CCD camera equipped with an objective lens. Individual video frames were exported at original resolution for subsequent color analysis using ImageJ, an open-source image analysis software. Color was examined in the red-green-blue (RGB) color space, in which each of the R, G and B channel reads a value from 0 to 255 for any chosen pixel.

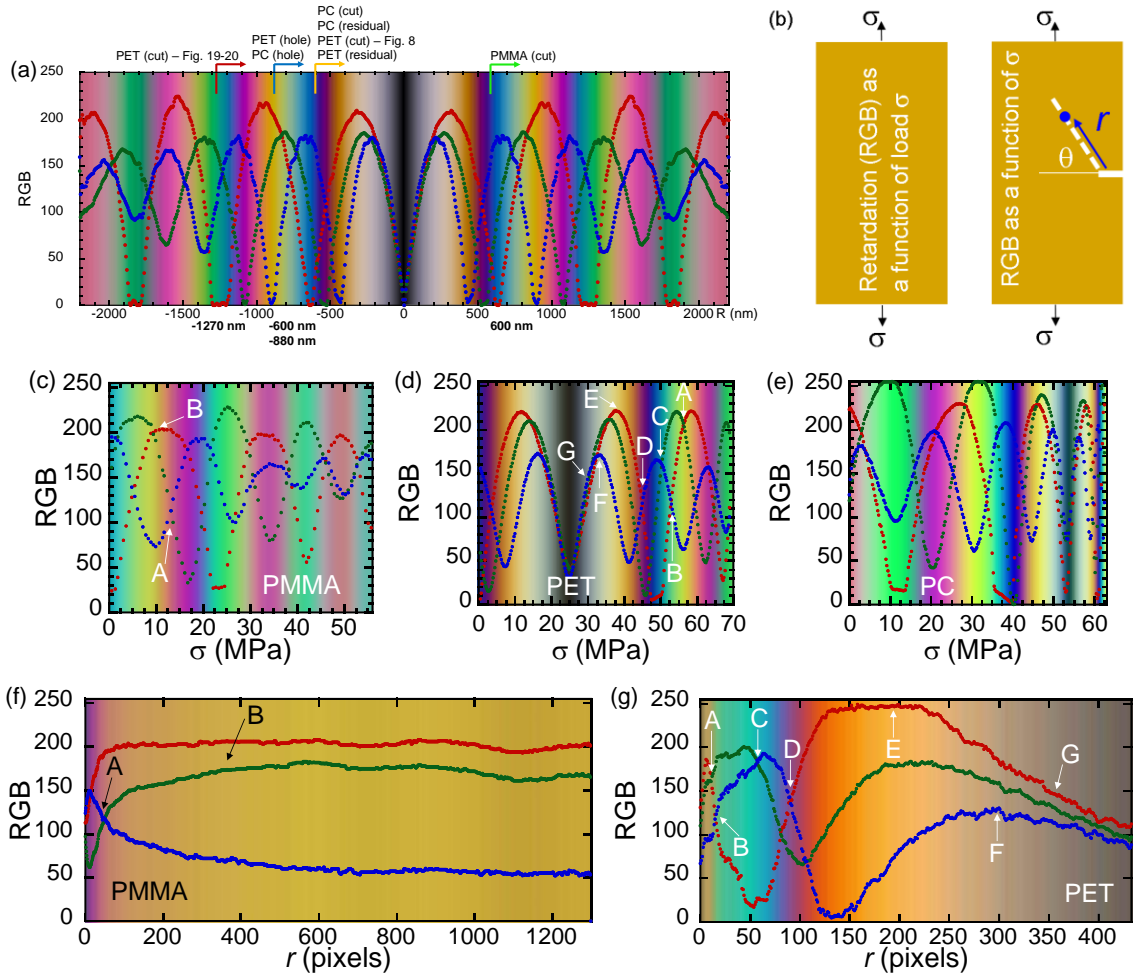

**Fig. 5.** (a) RGB profile of Michel-Lévy chart that can be used to quantify the birefringence changes during tensile drawing. The maxima, minima, and intersections of RGB were used as key features for comparison of the observed birefringence between uncut and precut specimens. Various negative retardation plates were employed to place various specimens at the initial retardation values as labeled by the vertical lines so that the retardation moves rightward during tensile drawing for all specimens. Retardation plates contributed in an additive (PMMA) or subtractive (PET, PC) manner to the initial emergence of birefringence so that the observed retardation passes zero for PET and PC specimens. (b) Schematic illustration to show that uncut specimens were drawn to record the emergent RGB as a function of stress and precut specimens drawn to record local RGB as a function of distance $r$ from the notch tip. (c)-(e) RGB in ML chart-like form is presented as a function of nominal load σ, obtained from respective drawing of uncut PMMA, PET and PC. Specifically, RGB variation in (c) is based on uncut stripe PMMA specimen ($L_0 \times W_0 \times H_0 = 50 \times 20.83 \times 5.5$ mm$^3$) drawn at $V/L_0 = 0.2$ min$^{-1}$. The RGB images in (d) and (e) are based on uncut PET and PC under drawing conditions described in Fig. 6a. (f) RGB variation as a function of distance $r$ from the notch tip at load σ = 11.4 MPa for PMMA specimen that is described in Fig. 7a-c. (g) RGB variation as a function of $r$ at σ = 15 MPa for PET that is described in Fig. 8a-c.

To obtain the stress-optical relationship (SOR) for each polymer, the RGB values were recorded during drawing of uncut specimen, correlating between RGB and stress (cf. Fig. 5b), as shown in Fig. 5c-e. The color images in Fig. 5c-e were generated using MATLAB. To determine the local stress in notched specimens, the RGB variation along a line drawn from the notch tip, as shown in Fig. 5b, was compared to the images in Fig. 5c-e for PMMA, PET and PC respectively. Since color is influenced by the choice of light and camera, each specimen was compared to images in Fig. 5c-e, generated based on the same combination of light source and camera. The locations of maxima, minima, and intersections of RGB values provided straightforward identification of stress. Using PMMA as an example, at location A in Fig. 5f the GB intersection in the second order is readily discerned from the same feature in the SOR, marked A in Fig. 5c. At low loads where RGB variation are less distinctive, we approximate the local stress based on the approximate ranking and trends of the RGB curves. For example, at location B in Fig. 5f the RGB are somewhat invariable over the rest of the distance from the notch tip. The red and green curves are close in value, with red atop the green curve. Since green was steadily increasing towards red, the B value in Fig. 5c may be an adequate estimate of B in Fig. 5f. The error introduced to the assignment of local stress by this pseudo-quantitative assignment is on the order of 0.4 to 2 MPa, corresponding to an uncertainty in the measurement of retardance on the order of 10 to 50 nm.

For PET and PC, both displaying significant birefringence, we used various negative retardation plates to record the birefringence, as labeled in Fig. 5a. Instead of constructing a new RGB vs. stress curve for each retardation plate combination that we used, we also prepared a stress vs. retardation plot from the data in Fig. 5d-e, as shown in Fig. 6a, by comparing the locations of the RGB intersections, maxima, and minima to those extracted from the ML chart of Fig. 5a. For example, for PET near the elastic-yielding transition (EYT), the intersections, maxima, and positions marked A-G in Fig. 5g were matched to similar features marked A-G in Fig. 5d.

Accurate determination of local stress as a function of distance $r$ from the notch tip requires the correct identification of the notch edge. With 4K CCD camera attached to a microscope objective lens a resolution on the range of $r_p = 1-4$ μm per pixel may be achieved. However, because of imperfections due to the cutting procedure, the polymers' mechanical response to the cutting, and slight misalignment of the camera relative to the cut opening, the notch edge usually appears blurry. This limits the available spatial resolution to, at best, $r_b \sim 40-80$ μm for the thick PMMA specimens, and 5-40 μm for PET and PC sheets. In addition, the above factors can lead to asymmetric gourd-shape patterns (fringes). In such specimens showing asymmetric color evolution during tensile drawing, we rely on the clearer half of the tip region for analysis.

**4. Experimental results**

According to the theoretical analysis and discussion in Section 2, for transparent plastics, e.g., amorphous glassy polymers, some key questions may be addressed using birefringence observation. For example, based on the SOR, we may map out the stress field near an intentional through-cut or circular hole as a function of the far-field stress, i.e., the nominal tensile stress that arises in response to the applied extension. The spatial variation of the stress field should distinguish the three scenarios sketched in Figs. 1a-c, with Fig. 1a representing the conventional picture. The birefringence observations may answer whether the stress field converges to the same stress level at different loads as shown in Fig. 1a or builds up sharply at the crack tip even at low loads as shown in Fig. 1b or attains different levels at the tip for the different loads as shown in Fig. 1c. In particular, we may be able to compare the tip stress with fracture or yielding strength ($\sigma_b$ or $\sigma_y$) determined from uncut specimens of the same material.

*4.1 Stress/strain optical relation and residual birefringence*

In glassy polymers significant birefringence can emerge during linear elastic deformation because backbone and side-group orientations can hardly relax. We exploit the one-to-one correspondence between birefringence and macroscopic stress, i.e., the SOR, using the setup shown in Fig. 3 for uncut specimens. Based on the Michel-Lévy chart, we establish SOR for PET and PC. SOR will allow us to quantify the local stress state during drawing of precut samples. Since PMMA shows little strain-induced birefringence whereas PET and PC are strongly birefringent, we employ retardation plates to displace the initial reading so that at load levels of interest the retardation changes are visibly more discernible, as indicated in Fig. 5a. The SOR for PET and PC is presented in Fig. 6a, along with the stress vs. strain curves in Fig. 6b, revealing a stress-optical relation $C = \sigma/|\Delta n|$. In agreement with the literature, the birefringence of PET and PC is about 17 times the C for PMMA. Since some experiments involved precut PET with annealing (to remove residual birefringence), Fig. 6a also provides SOR from annealed PET (PET-ann) for mapping of local stress field in such specimens.

Since PC is ductile at room temperature, significant residual birefringence may be expected before the shear yielding produces any visible necking. The photos in Fig. 6c show, for example, the evidence that irreversible, i.e., plastic deformation, has taken place at a nominal stress level (58 MPa) close to the yield stress $\sigma_y$. Similarly, residual birefringence may be expected in precut specimens upon sufficient loading and unloading.

*4.2 Birefringence observations of precut PMMA*

With an intentional through-cut, PMMA is a good candidate to explore the fracture behavior of brittle polymers. Berry's measurement of toughness of PMMA sixty years ago[1] has indicated that there is a relatively constant $G_{Ic}$ to describe how precut PMMA becomes progressively weaker with increasing cut size $a$. With birefringence observations, the fracture characteristics can be compared with the LEFM description, as discussed in Section 2.6.

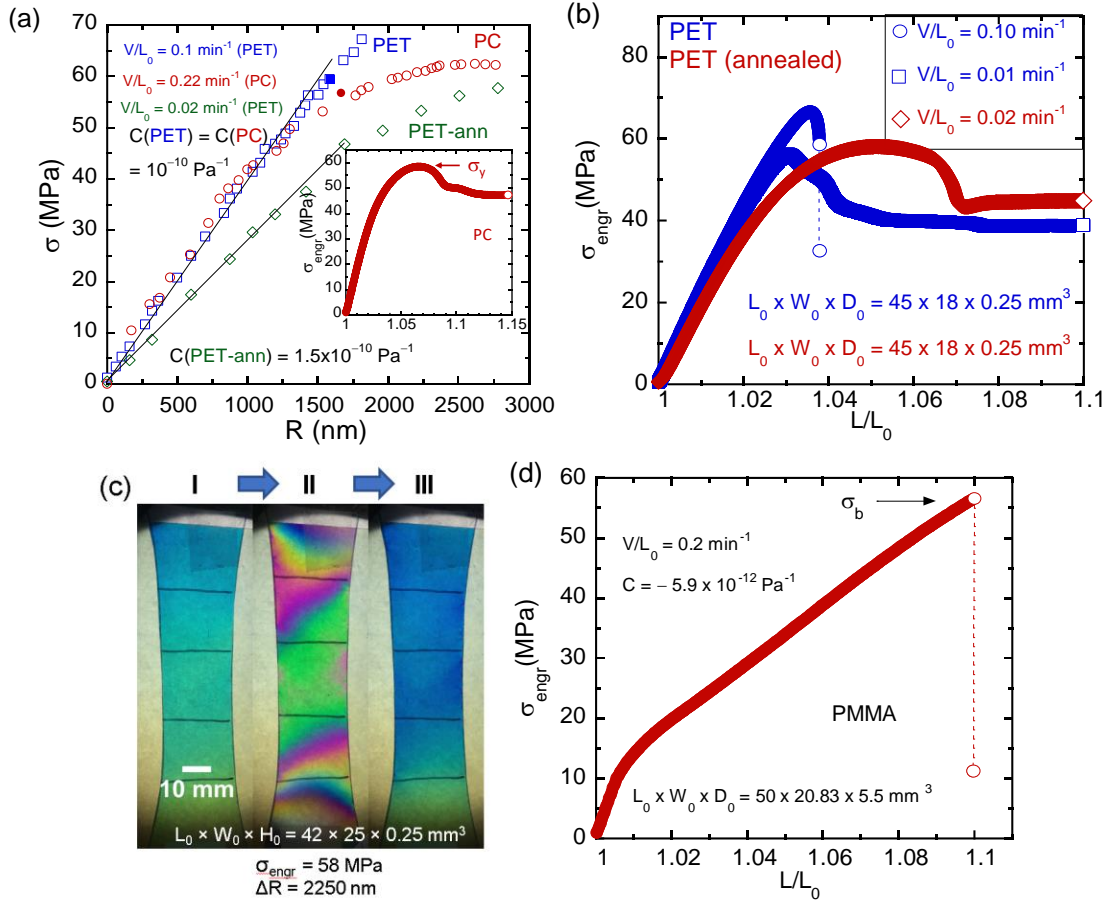

**Fig. 6.** (a) Stress-optical relationship (SOR) of PET (□ as-received; ◊ annealed for 2 hours at 75 °C) and PC (○ as-received) from tensile drawing of dogbone-shaped specimens. The stress-strain curve of PC is presented in the inset. (b) Stress-strain curves of PET at two different rates. The SOR in Fig. 6a was constructed from the as-received (○) specimen drawn at $V/L_0 = 0.10$ min$^{-1}$ and annealed specimen (◊) drawn at $V/L_0 = 0.02$ min$^{-1}$. (c) The onset of yielding in PC (shown here) at 58 MPa and PET (not-shown here) was confirmed by the appearance of residual birefringence in the specimen (III) as compared to the specimen prior to drawing (I). No discernable change in retardance before and after drawing was observed at $\sigma_{engr} < 50$ MPa for PC and $\sigma_{engr} < 60$ MPa for PET. (d) Stress-strain curve of uncut stripe PMMA corresponding to Fig. 5c.

PMMA is brittle, as shown by the SS curve in Fig. 6d. As expected from LEFM, PMMA becomes weaker upon introduction of a through-cut. The stress buildup at the cut tip can be quantified using the birefringence observation, based on the setup shown in Fig. 3. Fig. 7a contains a collection of snapshots at different moments from the video recording of a precut PMMA being drawn until fracture. Given the SOR in Fig. 5c showing a correspondence between RGB and stress, we report the actual stress field $T = \Delta n/C$ around the tip at $\theta = \pi/3$ by plotting T against $r^{-1/2}$, as shown in Fig. 7b. The intercepts reveal the stress level in the far-field, $\sigma_\infty$, which accurately matches $\sigma_0$. At $P^{-1/2} = 2.7$ mm$^{-1/2}$, i.e., $r = P = 0.14$ mm, we observe T to level off to $\sigma_{tip} = T(r = P)$. To reiterate, the $r$ dependence of $T(r)$ clearly reveals the form of Eq. (14d). Specifically, we can evaluate $K_{I(exp)}$ according to Eq. (14d) from the slopes of the straight lines in Fig. 7b where $\sigma_{tip}$ denotes the level of stress plateau in Fig. 7b. Fig. 7c shows that $K_{I(exp)}$ is indeed linearly proportional to the operational definition $K_I$ of Eq. (1). However, unlike a previous photoelastic measurement of $K_{I(exp)}$ for polysulfone[31] that found $K_{I(exp)} = K_I$ to approximately hold, we find $K_{I(exp)}$ to be only half of $K_I$. Plausibly, the deviation occurs because the cut size is small, not overwhelmingly larger than P. Plotting $\sigma_{tip}$ from Fig. 7b against $K_I$ of Eq. (1) in Fig. 7d reveals a linear relationship, between them, the combination of Eq. (14c-d) as

$$\sigma_{tip} = T(r = P) = \Delta n(r = P)/C = QK_I, \text{ with}$$
$$Q = [1/(\pi a)^{1/2} + (3^{1/2}/2)(K_{I(exp)}/K_I)/(2\pi P)^{1/2}]. \quad (16)$$

Specifically, given $a = 1.1$ mm, $K_{I(exp)}/K_I \sim 0.5$, we have $P = 0.11$ mm from the slope Q in in Fig. 7d. In the limit of $a \gg P$ we can expect $Q \to (3^{1/2}/2)/(2\pi P)^{1/2}$. Such linearity between $\sigma_{tip}$ and $K_I$ in fact describes the trend of a master curve based on similar data involving five different cut sizes, as shown in Fig. 7e. The slope in Fig. 7e is Q, revealing the value of P that marks the onset of the stress saturation (SS). However, since PMMA does not yield, the emergence of the SS may not be taken as a sign of plastic zone formation. In the other words, the emergence of the SS zone may not be tip blunting due to plastic deformation and may be unrelated to HRR[56, 57] singularity. More discussion is deferred to Section 5.1.

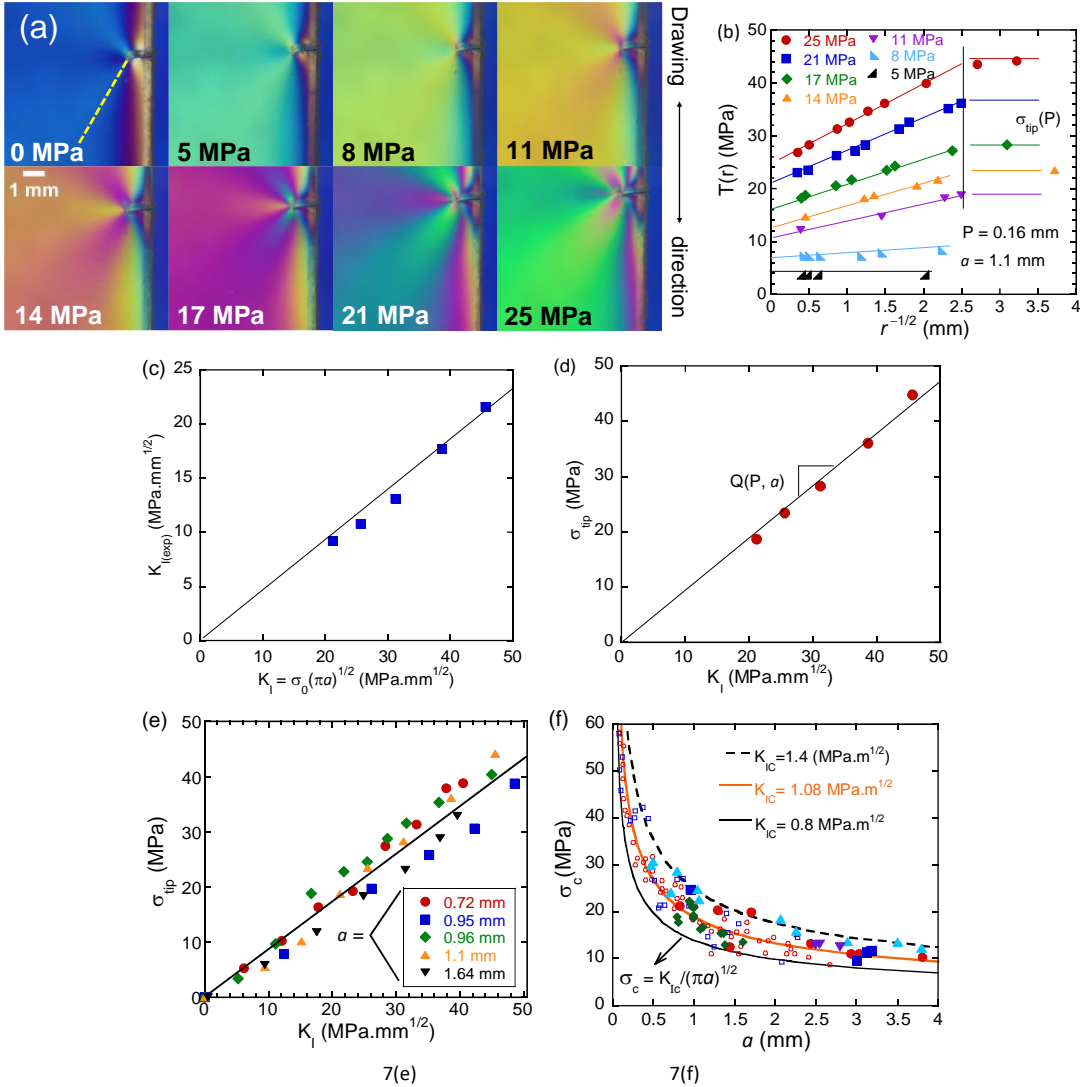

**Fig. 7.** (a) Stripe PMMA specimen ($L_0 \times W_0 \times D_0 = 130 \times 21.4 \times 5.5$ mm$^3$) with B-type cut size $a = 1.1$ mm was drawn at $V/L_0 = 0.077$ min$^{-1}$ to failure. The RGB profile along the diagonal dashed line at each load $\sigma_0$ (0 to 25 MPa) was compared to Fig. 5c to obtain the plot in Fig. 7b. (b) Local tensile stress $T = (\sigma_{yy} - \sigma_{xx})$ plotted against $r^{-1/2}$ at different stages during drawing that produces images in (a), showing stress saturation at high values of $r^{-1/2}$. (c) $K_{I(exp)}$ in (3a) evaluated from the slopes in (b), plotted against the operational definition of $K_I$ in Eq. (1). (d) Tip stress read from the stress plateau in (b) plotted against $K_I = \sigma_0(\pi a)^{1/2}$. (e) Tip stress involving B-type notch as a function of $K_I$. The data fits to a straight line, giving rise to Eq. (16). (f) Berry style plot showing variation of $\sigma_c$ with cut size $a$, involving Berry's data[1] (○ $D_0 = 4.8$ mm and □ $D_0 = 1.6$ mm) and our data for PMMA. The effect of different cut types in our PMMA data are denoted by the different symbols: ● B-type ($D_0 = 3.0$ and 5.5 mm), ■ C-type ($D_0 = 2.5$ mm), ♦ R-type (sharp) ($D_0 = 2.5, 3.0,$ and 5.5 mm), ▲ R-type (blunt) ($D_0 = 5.5$ mm), and ▼ MR-type cuts ($D_0 = 2.5$ mm).

The data spread around the straight line in Fig. 7e indicate a variation in P. This spread can also be observed in a Berry style plot. For example, plotting the present data along with Berry's data[1], we see a spread in Fig. 7f, suggesting $K_{Ic}$ is not strictly a constant. As indicated in Fig. 7f, $K_{Ic}$ varies from 0.8 to 1.4 MPa.m$^{1/2}$. We defer to Section 5 our further discussion on the origin of these spreads in Figs. 7e-f.

*4.3 Birefringence observations of precut PET and PC*

"If toughness is sufficiently high, fracture mechanics ceases to be relevant to the problem because failure stress is insensitive to toughness…".[38] While high toughness here might mean that inherent flaws do not dictate failure behavior of uncut specimens, in presence of large precut, LEFM can still provide the intellectual guidance. For example, the idea of stress intensification can still be expected to apply before yielding, and the concept of plastic zone formation may be more readily explored. In the regime of LEFM, i.e., before the onset of yielding, we should be able to find out whether the local stress builds up near the crack tip according to Eq. (3a). More importantly, we may determine using SOR and birefringence measurements whether the limiting stress at the tip, to be taken as $\sigma_{Y(inh)}$, is comparable to or much higher than $\sigma_y$.

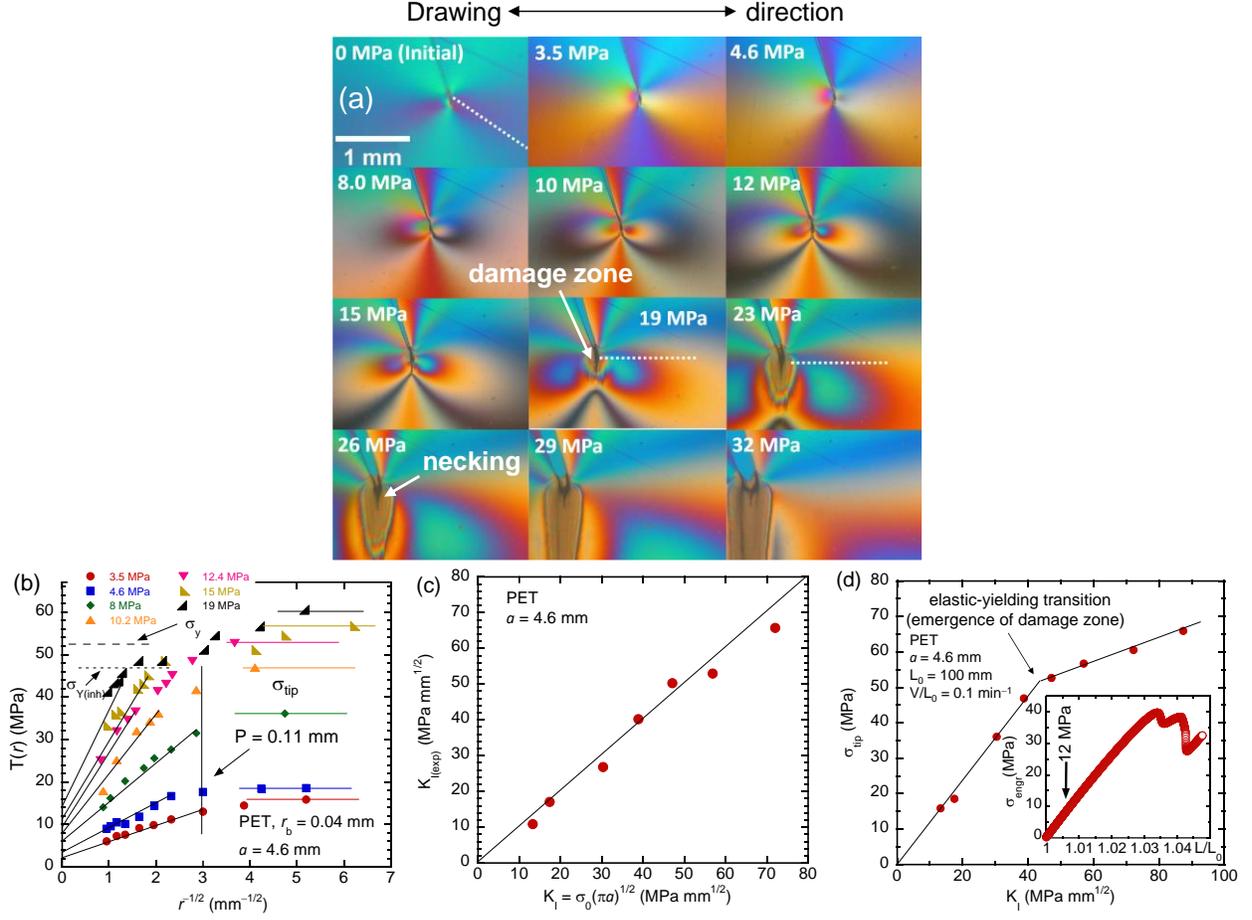

**Fig. 8.** (a) Stripe PET specimen ($L_0 \times W_0 \times D_0 = 100 \times 42 \times 0.25$ mm$^3$) with B-type cut of size $a = 4.6$ mm was drawn at $V/L_0 = 0.10$ min$^{-1}$. The RGB profile along the diagonal dashed line at each load (aside from a different line at $\sigma_0 = 19$ and 23 MPa) was compared to Fig 5a and 6a to obtain the plot in Fig. 8b. (b) Local tensile stress T plotted against $r^{-1/2}$ at different stages during drawing that produces images in (a). (c) $K_{I(exp)}$ in Eq. (3a) evaluated from the slopes in (b), plotted against the operational definition of $K_I$ in Eq. (1). (d) Tip stress read from the stress plateau in (b) plotted against $K_I = \sigma_0(\pi a)^{1/2}$ where the inset shows the stress vs. strain curve of the precut specimen.

In presence of a precut, PET sheets behave in a ductile manner. As the specimen is drawn, the birefringence and therefore stress build up at the cut tip until a damage zone emerges at $\sigma_0 = 19$ MPa, followed by growth of the damage zone (e.g., at 23 MPa), necking (at 26 MPa) and subsequent crack advancement (after 32 MPa), as shown by the photos in Fig. 8a. Quantifying the local stress using the SOR in Fig. 5d to convert from RGB profile along the dashed line ($\theta=\pi/3$) in Fig. 8a to T, i.e., following Eq. (14c), we report the local tensile stress field in Fig. 8b as a function of distance $r$ from the cut tip, with a spatial resolution of 0.04 mm. Like Fig. 7b for brittle PMMA, Fig. 8b also confirms stress buildup as $r^{-1/2}$, shown by the straight lines, which are drawn using the far field stress $\sigma_0$ at different stages of tensile extension as the intercept. Since yielding at the tip did not occur up to $\sigma_0 = 10$ MPa, the stress plateau (involving the lowest $\sigma_0$ from 3.5 to 10.2 MPa) shares the same physics that produces the same characteristics in Fig. 7b for PMMA. We should also note that the onset of the stress plateau (or saturation) is at a scale of $r^{-1/2} = 3.0$ mm$^{-1/2}$, comparable to the location observed of PMMA.

For $\sigma_0 > 10$ MPa, the tip stress reaches inherent yield strength $\sigma_{Y(inh)}$, well before $r$ reaches P, and saturates because of plastic deformation. Thus, the emergent stress plateaus in Fig. 8b should be viewed to arise from two different causes. Moreover, we find $\sigma_{Y(inh)}$ to be comparable to $\sigma_y$ from Fig. 6b, as indicated by the two horizontal dashed lines.

From the slopes indicated by the straight lines in Fig. 8b we can evaluate $K_{I(exp)}$ using Eq. (3a). Fig. 8c shows $K_{I(exp)}$ to (1) linearly increase with $\sigma_0$ or $K_I$ of Eq. (1) and (2) to be comparable to $K_I$. Thus, contrasting Fig. 7c, Fig. 8c implies that $a \gg P$. In this limit, we can expect $K_I$ to take on the theoretical result of Eq. (1). Since both PMMA and PET have comparable P, the difference comes from that in the cut size, $a = 1.1$ mm for PMMA and 4.6 mm for PET. For PET, the condition of $a \gg P$ is more strongly satisfied, so that $Q \sim (3^{1/2}/2)/(2\pi P)^{1/2}$ in Eq. (16). Fitting the slope in Fig. 8d to Eq. (16), we find $P = 0.12$ mm in agreement with the onset of stress saturation at 0.11 mm in Fig. 8b.

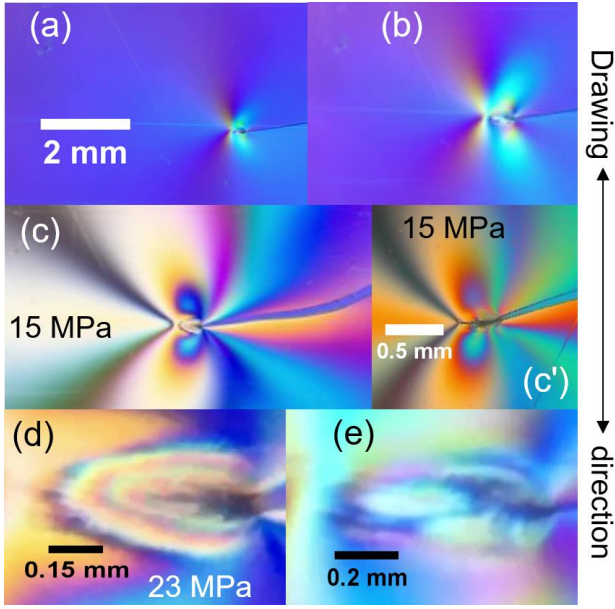

**Fig. 9.** Precut PET specimen ($L_0 \times W_0 \times D_0 = 100 \times 41 \times 0.25$ mm$^3$) with B-type (cut in LN2) cut of $a = 9.4$ mm in (a) undeformed state and (b) after unloading from applied nominal load $\sigma_0 = 15$ MPa (at $V/L_0 = 0.10$ min$^{-1}$) that produces the retardance pattern in (c), in comparison with (c'), which is the same image from Fig. 8a at 15 MPa. The difference in coloration arises from the use of different video cameras; the light source was kept the same between specimens. (d) Magnification of image (c) shows the emergence of a damaged zone (pear-shaped) in front of the crack tip. (e) Magnification of image (b), whose size is similar to that of the pear-shaped damaged zone (d), shows that the damaged zone at the crack tip involves irreversible deformation.

It is also interesting to note that $K_{I(exp)}$ starts to decrease after 40 MPa.mm$^{1/2}$, which is a sign of yielding at the cut tip. Upon further drawing, PET undergoes considerable plastic deformation in the form of necking as shown by the images (26 MPa and higher) in Fig. 8a. A transition from elastic extension to yielding (EYT), absent in PMMA, can be explicitly identified by plotting $\sigma_{tip}$ against $K_I$. As shown by Fig. 8d, the EYT occurs at the kink around 40 MPa.mm$^{1/2}$, which is indiscernible at 12 MPa in the stress vs. strain curve, shown as the inset of Fig. 8d. Thus, for both PET and PC, $K_{Iy}$ at YET hardly captures the overall mechanical response, as the nominal load triples during necking from the cut tip before the eventual macroscopic separation.

Beyond 12 MPa, i.e., above EYT, a damage zone becomes observable. In fact, this EYT can be verified in terms of residual birefringence as follows. In Figs. 9a-c, we show with a separate precut PET specimen that upon loading to a far-field stress level of 15 MPa, whose image in Fig. 9c corresponds well to that labeled by 15 MPa in Fig. 8a and reproduced as Fig.9c', the local stress has exceeded the level for yielding: Upon unloading from this state (Fig. 9c), the retardation shows a visible change in Fig. 9b over the image taken of the initial undeformed state in Fig. 9a. The appearance of residual birefringence confirms that there is yielding and plastic deformation at the stress level indicated by Fig. 9c. We recognize the state at 23 MPa in Fig. 8a as revealing damage because upon unloading from such a state, as shown in Fig. 9d, the material in the damage zone exhibits strong residual birefringence in Fig. 9e.

PC is ductile in its uncut form as shown by the stress vs. strain curve in Fig. 6a. To assess the generality of the phenomenology revealed by PET in Fig. 8a-d and Fig. 9, it is necessary to examine another case based on PC. The snapshots in Fig. 10a from the video recording of a precut PC specimen under uniaxial drawing indeed reveal similar features, such as the butterfly-like retardation patterns around the cut tip, emergence and growth of the damage zone, necking and subsequent crack propagation. Notably, the region of stress intensification keeps growing with rising load. A quantitative analysis of these images produces Fig. 10b, confirming that the local stress buildup at the tip also does not approach the limiting value at low loads. On the other hand, the stress buildup shows pronounced slowing-down after 40

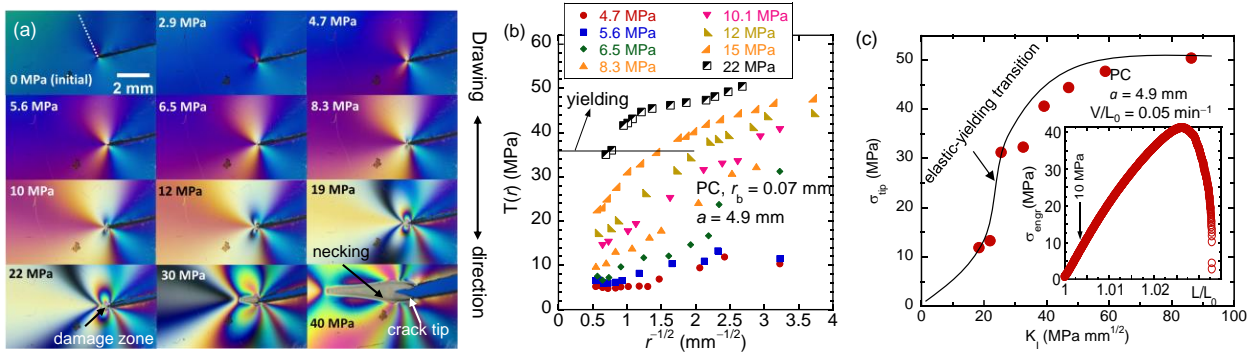

**Fig. 10.** (a) PC specimen ($L_0 \times W_0 \times D_0 = 100 \times 36 \times 0.25$ mm$^3$) with B-type cut size $a = 4.9$ mm was annealed at 160 °C for 1 hour under constraint prior to tensile extension to remove the residual birefringence leftover from cutting. The specimen was drawn at $V/L_0 = 0.05$ min$^{-1}$ to failure. The RGB profile along the diagonal dashed line at each load was converted using Fig 5a and 6b to provide data in (b). (b) Variation of local tensile stress T with distance $r$ from the notch tip. (c) Increase of the stress level at the notch tip $\sigma_{tip}$ with the applied load. At 5 MPa, local stress drastically increases, marking the elastic-yielding transition. The inset shows the stress-strain behavior of the precut specimen, with the appearance of the local damaged zone and necking occurring well before the nominal stress maximum.

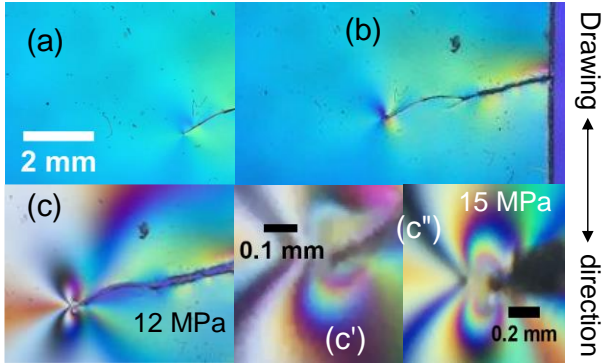

**Fig. 11.** PC specimen ($L_0 \times W_0 \times D_0 = 100 \times 28 \times 0.25$ mm$^2$) with B-type cut size $a = 6.4$ mm in (a) its undeformed state and (b) after unloading from applied nominal load $\sigma_0 = 12$ MPa (at $V/L_0 = 0.05$ min$^{-1}$) that produces the retardance pattern in (c), in comparison with (c") the PC from Fig. 10a at $\sigma = 15$ MPa. (c') Magnification of image (c) shows that the damaged zone at the crack tip involves irreversible deformation. (a-c) The PC specimen was annealed at 160 °C for 1 hour under constraint prior to tensile extension to remove the residual birefringence leftover from cutting.

MPa. When examining the tip stress $\sigma_{tip}$ as a function of load $\sigma_0$, Fig. 10c reveals a rather sharp EYT. In contrast to PET that shows a gradual increase of $\sigma_{tip}$ with $\sigma_0$, the approach to yielding is steep. On the other hand, similar to PET, no EYT can be detected based on the stress vs. strain curve in the insert of Fig. 10c.

We can also show with separate precut PC specimen in Fig. 11a-c how residual birefringence is produced at a far-field stress level of 12 MPa. After this load that produces the retardation distribution shown in Fig. 11c, unloading leads to a different image in Fig. 11b from that in Fig. 11a, which is evidence for residual birefringence. Here a magnification of the retardation distribution at the cut tip in Fig. 11c resembles the state at 15 MPa from PC in Fig. 10a, shown as Fig. 11c".

In conclusion, based on brittle PMMA as well as PET and PC, we find Fig. 1c to be a more realistic depiction of the stress intensification. More importantly, the quantitative birefringence observations suggest that the inherent strength $\sigma_{F(inh)}$ or $\sigma_{Y(inh)}$ is comparable to, in fact lower than the brittle or yield stress, i.e., $\sigma_b$ for PMMA or $\sigma_y$ for PET and PC respectively. As expected, the ductile responses of PET and PC occur in the form of yielding and plastic deformation while PMMA shows no sign of any EYT.

Before EYT, according to Fig. 8d for PET and Fig. 10c for PC, $\sigma_{tip}$ also increases approximately linearly with $K_I$, as shown in Fig. 12a. Before EYT, i.e., below 60 MPa, the data look similar to PMMA data in Fig. 7e, i.e., approximately following Eq. (16). According to Eq. (3a), $\sigma_{tip}/K_I$ should be constant, independent of $K_I$. In other words, according to the stress intensification analysis for SEN, the stress intensification, given by $\sigma_{tip}/\sigma$, should not depend on the load level $\sigma_0$. This description can be scrutinized for different cut sizes. Replotting Fig. 12a as Fig. 12b, we show that for PMMA, PET and PC before EYT the stress intensification reveals a tendency to increase with $K_I$. According to Eq. (16), the monotonic rise of $\sigma_{tip}/K_I$ may correspond to a decrease in P whereas the decline from the local maximum or plateauing marks the onset of EYT: The tip stress ceases to build as fast as the load upon yielding. PMMA does not show such a cusp because it does not undergo yield prior to brittle fracture. PMMA, given in open symbols, shows more gradual change, with $\sigma_{tip}/K_I$ confined between 0.60 to 1.2 mm$^{-1/2}$, in part because $a$ is much smaller. In contrast, PET and PC indicate a significant increase of $\sigma_{tip}/K_I$ from 0.20 to 1.60 mm$^{-1/2}$, with $\sigma_{tip}/\sigma_0$ reaching up to 9.0.

*4.4 Birefringence measurements of PET and PC with circular hole*

The challenge to establish the findings presented in the preceding Section 4.2 for brittle PMMA and 4.3 for ductile PET and PC resides in whether we can claim to have sufficiently high spatial resolution. The stress field quantified through the birefringence observations is also sensitive to the characteristics of the intentional through-cut such as sharpness that is not trivial to quantify. In contrast, smooth circular cracks are relatively easy to introduce and simple to quantify in terms of the stress intensification, as discussed in Section 2.8. Therefore, we investigate the stress buildup at a circular through-crack till failure, signified by visible strain localization (necking) at the poles.

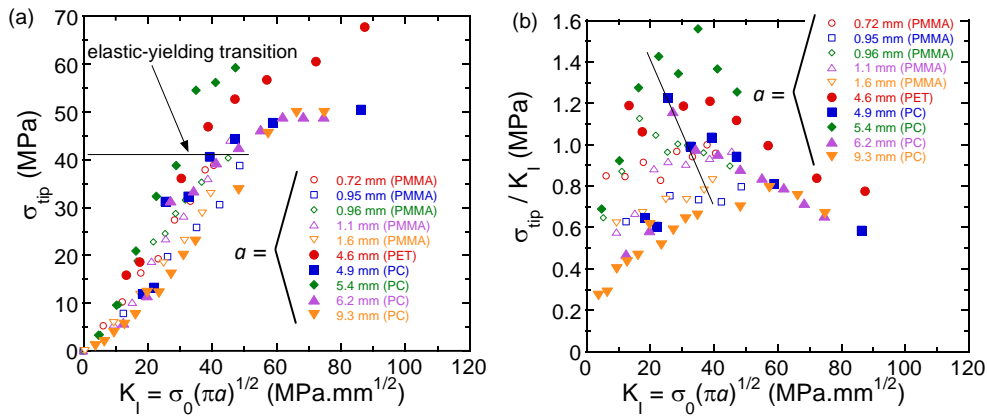

**Fig. 12.** (a) Evolution of the stress observed at the notch tip as a function of the load characterized by $K_I$, including the PMMA data from Fig. 7e, along with the PET data from Fig. 8 and PC data with from Fig. 10. The remaining PC specimens have B-type notch. The EYT of PC from Fig. 10 and onset of necking at tip for PET from Fig. 8 are denoted. (b) Ratio of $\sigma_{tip}$ and $K_I$ shows how the characteristic length scale P, identified in Eq. (16) varies during drawing.

Fig. 13a consists of a collection of snapshots from video recording of the retardation field produced around a large circular hole in PET during tensile drawing. Using the SOR, these images can be quantitatively converted to describe the stress distribution as a function of distance $r$ from the center of the hole, as shown in Fig. 13b in terms of the normalized local stress. Approximately, before yielding the stress field is proportional to the load level because all the curves nearly collapse onto one curve. Rewriting the LEFM prescription (Eq. 15) for the stress along the direction perpendicular to the drawing direction in normalized form, we have

$$\Omega(r) = \sigma_{zz}(r, \theta=\pi/2)/\sigma = 1 + 0.5(a/r)^2 + 1.5(a/r)^4, \quad (17)$$

and the experimental data are lower in magnitude than $\Omega$ (the smooth curve), as shown in Fig. 13b. For example, the stress at the two poles is always less than three times the loading level. Thus, we conclude that the stress at the pole before necking at 32 MPa, equal to 2.24×32 MP = 72 MPa is comparable to $\sigma_y \sim 60\text{-}70$ MPa, as shown in Fig. 6b.

A similar study on PC reveals the analogous phenomenology, as shown in Fig. 14a-b. In other words, the quantitative birefringence observations of PC with circular hole also confirm the conclusion based on PET with SEN that local yielding and necking at the edge of the circular crack take place before necking at 36 MPa, equal to 1.75×36 = 63 MPa, at a stress level comparable to $\sigma_y = 60$ MPa. Another common feature in comparison between Fig. 13b and Fig. 14b is that in both cases the stresses cease to increase to the edge of the hole as strongly as shown by $\Omega$ function in Eq. (17) at high loads. This suggests the emergence of yielding.

According to Fig. 6a the SOR is linear for PET and PC up to a retardation level of ca. 1500 nm. Beyond this level, yielding starts to take place, as indicated in Fig. 6a-b, evidenced by residual birefringence shown in Fig. 6c. Consistent with this information, Fig. 15a reveals linear relationship between retardation at the two poles, $R_{pole}$ and load $\sigma$ up to ca. $R_{pole} = 1.9$ μm, implying that the local stress $\sigma_{zz}$ is linear in $\sigma$, as can be expected from the LEFM formula, Eq. (15). Figs. 13b and 14b indeed show that the normalized stress distribution collapse onto a single curve at low loads. At sufficiently high loads, at small vales of $a/r$ where retardation is low, i.e., away from the circular hole, the data still approximately overlap. However, at $a/r \sim 1$, i.e., close to the edge of the circular holes, splitting occurs, which is especially evident in Fig. 15b because PC is more ductile. The occurrence of yielding is readily evident when the retardation at the poles is plotted against the load as shown in Fig. 15a. Upon yielding, the linear elasticity breaks down, and the retardation $R_{pole}$ at the poles rises faster than linearly with the applied load. In other words, interestingly, the upward deviation from the linearity in Fig. 15a may be regarded as a signature for yielding.

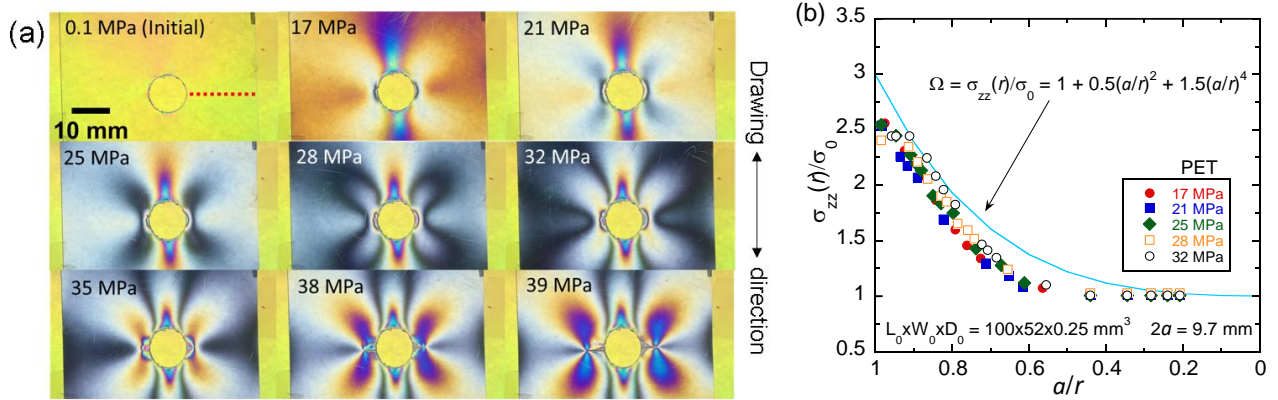

**Fig. 13.** (a) PET specimen ($L_0 \times W_0 \times D_0 = 100 \times 52 \times 0.25$ mm$^3$) with circular hole (through-crack) of diameter $2a = 9.7$ mm was drawn at $V/L_0 = 0.10$ min$^{-1}$ to failure. Necking was observed above 32 MPa. (b) Local stress $\sigma_{zz}$ variation along the horizontal dashed line in (a) was normalized by the applied load $\sigma_0$ to compare to Eq. (13) (solid line).

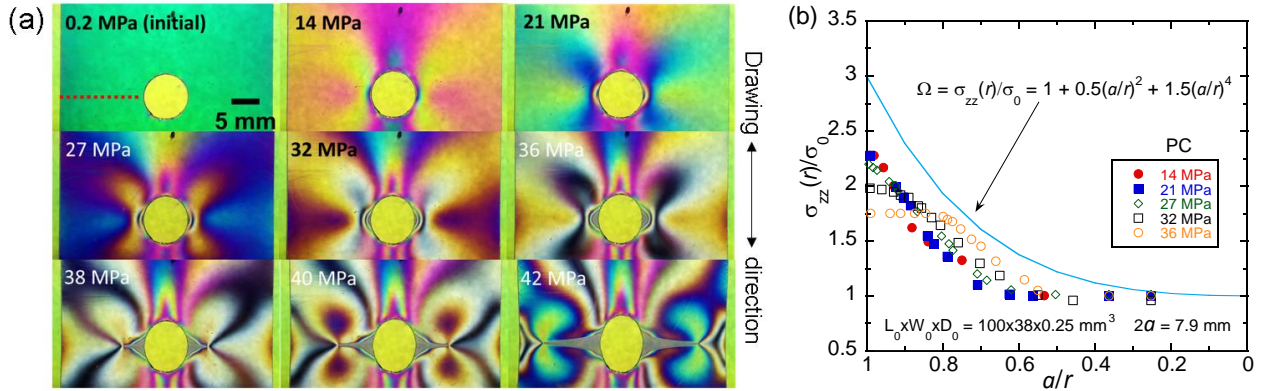

**Fig. 14.** (a) PC specimen ($L_0 \times W_0 \times D_0 = 100 \times 38 \times 0.25$ mm$^3$) with circular hole of diameter $2a = 7.9$ mm was drawn at $V/L_0 = 0.05$ min$^{-1}$ to failure. Necking was observed above 36 MPa. (b) Local stress $\sigma_{zz}$ variation along the horizontal dashed line in (a) was normalized by the applied load $\sigma_0$ to compare to Eq. (13) (solid line). The noticeable decline in $\sigma_{zz}/\sigma_0$ at higher applied load is due to the significant softening of PC during yield, as evidenced by the stress-strain curve inset in Fig. 6b.

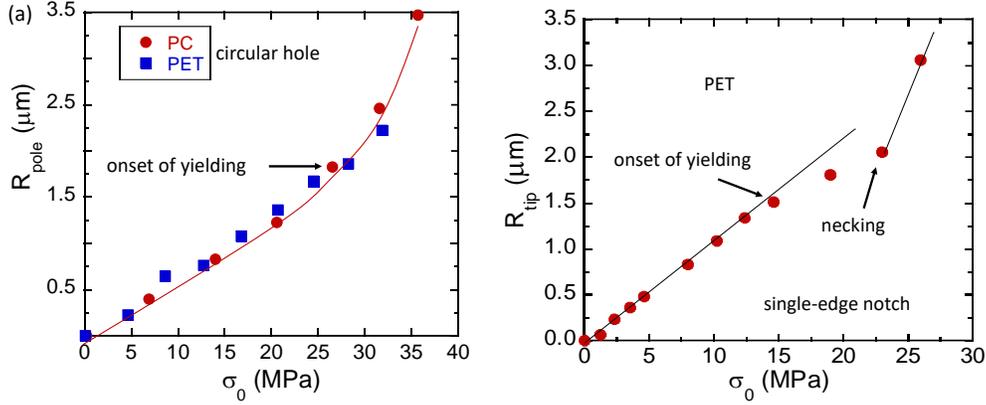

**Fig. 15.** (a) Retardation at the pole ($r = 1.01a$ to $1.05a$) of the circular crack against the applied load for PET from Fig. 13 and PC from Fig. 14. (b) Retardation at the notch tip against the applied load for PET based on data from Fig. 8.

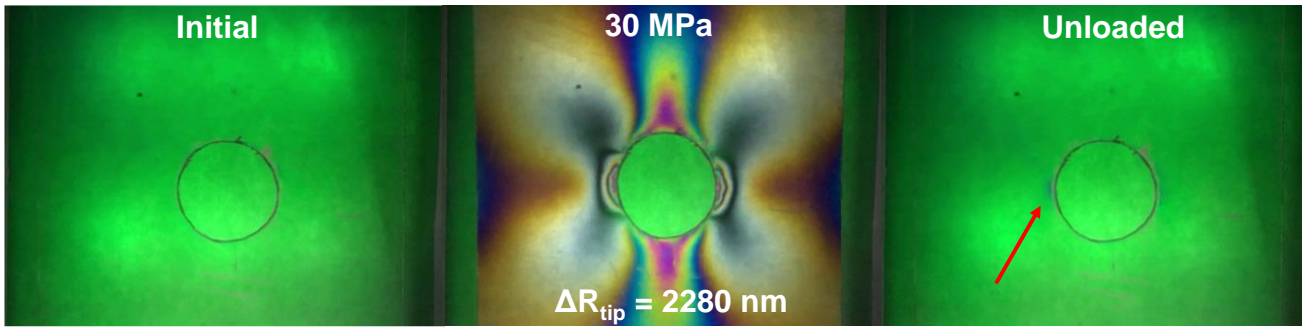

**Fig. 16.** Residual birefringence observed at the red arrow in PET stripe specimen ($L_0 \times W_0 \times D_0 = 80 \times 37 \times 0.25$ mm$^2$) with circular crack of diameter $2a = 9.7$ mm after drawing at $V/L_0 = 0.33$ min$^{-1}$ to $\sigma_{engr} = 30$ MPa and immediately unloading.

In contrast, at a through-cut tip, the retardation buildup weakens upon yielding, as shown in Fig. 15b. In other words, $R_{tip}$ hardly increases after 1.5 μm for PET, showing the opposite trend to that in Fig. 15a. This difference stems from that in the nature of the local strain field. SEN produces mode I deformation of tensile opening, the strain field at the poles of the circular hole allows the retardation to run away upon yielding.

The phenomenon of residual birefringence can also be directly demonstrated by drawing a hole-containing specimen until a high load to build up sufficient retardation at the poles, followed by unloading. For example, as shown in Fig. 16, PET with circular crack is drawn to a tensile stress of 30 MPa, producing $R_{pole} = 2280$ nm. Unloading leaves a discernibly different retardation distribution relative to the initial, undeformed state.

## 5. Discussions

This study aims at exploring the physics governing fracture mechanical behavior of glassy polymers. We presented our investigation by first presenting theoretical analysis in Section 2 and *in situ* birefringence experiments in Section 4, guided by the basic principles of fracture mechanics. Specifically, using optical birefringence based on a white light source (cf. Fig. 3) and Michel-Lévy chart (Fig. 5) we quantify the local stress field near crack tip within a spatial resolution of 0.02-0.08 mm. Our observations qualitatively confirm the stress intensification concept of LEFM. On the other hand, the stress buildup shows quantitative deviation from the theoretical description. In particular, the inherent strength $\sigma_{F(inh)}$ or $\sigma_{Y(inh)}$, regarded to equal the local stress at the through-cut tip, is comparable to the nominal failure stresses ($\sigma_b$ and $\sigma_y$) for brittle and ductile glassy polymers respectively. Thus, no flaws have reduced fracture or yield strength in uncut specimens if they are present at all. In other words, there is no need to perceive inherent flaws when account for fracture or failure behavior of glassy polymers. Instead of using Eq. (8) to describe fracture strength where $a^*$ is fictitious and the magnitude $G_{Ic}$ is unknown *a* *priori*, we can resort to a recent network picture[14] that explained why brittle fracture occurs and estimated the magnitude of $\sigma_{F(inh)}$.

In presence of a sizable crack, we need to resort to fracture mechanics that has successfully summarized the phenomenology. For example, there exists a Griffith[40]-Irwin[42] like critical (strain) energy balance rate $G_{Ic}$ or critical stress intensity factor $K_{Ic}$ for a given material with different crack sizes. Inspired by the Westergaard-Irwin's description[25, 26] of stress intensification, we set out to show what determines $K_c$ or $G_c$. The cut size dependence of the critical load, expressed through $G_{Ic}$ in Eq. (6), i.e., $\sigma_c \sim a^{-1/2}$, can be understood to reflect the fact that there is stress intensification in front of a crack, as shown by Eq. (3a). According to the second pillar of fracture mechanics, local stress at crack tip would always reaches $\sigma_{Y(inh)}$ in a plastic zone that grows according to Eq. (5) until fracture. Our birefringence measurements reveal in Sections 4.2 and 4.3 that before failure in the form of either brittle fracture or yielding and necking at lower loads the stress level at crack tip only gradually increases toward a level bounded by fracture strength $\sigma_b$ of uncut specimens. These measurements have instructed us to explore another layer of meaning for $K_{Ic}$ so that it can be estimated based on material properties.

*5.1 Meanings of $K_{Ic}$ and P*

At fracture, corresponding to the last points in Fig. 7d-e, we find $\sigma_{tip}(K_{Ic})$ = 40 MPa, which is lower than breaking stress $\sigma_b$ of uncut PMMA given in Fig. 6d. This tip stress should be comparable to the inherent fracture strength $\sigma_{F(inh)}$ under the plane strain condition and is lower than $\sigma_b$ (as expected) that is measured under plane stress condition. We further note that $\sigma_{tip}$ does not reach $\sigma_{F(inh)}$ at earlier stages of drawing, in contrast to the scenario of LEFM depicted in Fig. 1a.

Since Eq. (16) appear to hold all the way up to fracture, regarding $\sigma_{tip}(P, K_{Ic})$ as $\sigma_{F(inh)}$, we can rewrite it to reveal the deeper meaning of $K_{Ic}$

$$K_{Ic} = \sigma_{F(inh)}/Q(P, a) \rightarrow \sigma_{F(inh)}(2\pi P)^{1/2}, \qquad (18)$$

where and henceforth we drop the numerical prefactor $3^{1/2}/2$ to emphasize the simplicity of final expression: With $a \gg P$, Q in Eq. (16) is only a function of P. Comparing the limiting result of Eq. (18) with Eq. (2a), we have $r_p^* = P$. Given $\sigma_{F(inh)}$ is comparable to $\sigma_b$, we also find $a^* = 2P$. Furthermore, comparing Fig. 7b to the conjectured scenario of Fig. 1c, P = $r_{pl}$. In other words, P in Eqs. (16) and (18) is a characteristic length scale below which stress intensification ceases. We speculate that P depends on the material response during crack formation.

In other words, P in Eq. (18) may be related to the curvature of the opening introduced by precutting. The tip bluntness characterized by P assures that the stress would saturate according to theoretical analyses and finite-element calculations[58, 59]. Moreover, P might also be the spatial extent where the sample undergoes structural failure during crack propagation, as shown and discussed in Fig. 17 (to be introduced in Section 5.3). In other words, treating $K_{Ic}$ as a material parameter, as given in Eq. (18), where P may vary with the cut characteristics, we are finally able to estimate its magnitude. We can compute $G_{Ic}$ from $K_{Ic}$ using the

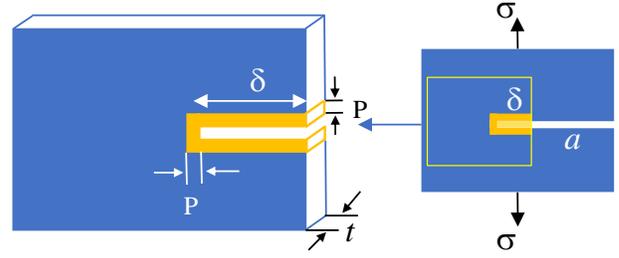

**Fig. 17.** Illustration of the concept of characteristic length scale P arising during crack propagation to form a new crack length of $(a + \delta)$ in a precut specimen of thickness *t*. The infected region of size $\delta$ is magnified in the left sketch where the yellow (color online) areas plausibly suffer structural failure.

classic relation Eq. (6) but cannot estimate $K_{Ic}$ from $G_{Ic}$. In other words, while Eq. (6) is valid, $K_{Ic} = (G_{Ic}E)^{1/2}$ has no meaning since we do not have an expression for $G_{Ic}$ except for its operational definition of $G_{Ic} = \pi a \sigma_c^2/E$, which circularly gives back the operational definition for $K_{Ic}$

In fracture mechanics $K_{Ic}$ is regarded as material constant. Berry's data show in Fig. 7f that $K_{Ic}$ varies over a range. Thus, $K_{Ic}$ is not rigorously a material constant but rather a material parameter in the sense that it explicitly depends on $\sigma_{F(inh)}$. The variation in $K_{Ic}$ arises from a variation in Q produced by different P since $\sigma_{F(inh)}$ can be regarded as a material constant, and since *a* dependence of Q in Eq. (16) can be expected to be weak.

It is worth noting that Eq. (16) can be rewritten to assume a form similar to Inglis' solution[38, 60] for stress concentration at the tip of an embedded elliptical opening

$$\sigma_{tip} = \sigma_0[1 + (3^{1/2}/2)(K_{I(exp)}/K_I)(a/2P)^{1/2}]. \qquad (19)$$

By comparison, we see P to be essentially related to the radius of curve of the opening. Thus, P is indeed a geometric parameter characterizing the cut sharpness. In the case of PMMA, the cuts are sufficiently blunt and small (cf. Fig. 7e) to the extent that Inglis solution is a much more realistic depiction of the stress intensification.

For a given cut size, our data in Fig. 7e and Fig. 12a can be described by Eq. (16). The observed linear relation simply means that the stress intensification due to a precut is essentially constant until fracture, as described by Eq. (19). At fracture, we can rewrite Eq. (19) as

$$\sigma_c = \sigma_{F(inh)} f(a/P) \qquad (20)$$

for the fracture strength of a precut specimen. This expression explicitly bypasses or omits the need to introduce $K_I$ and $K_{Ic}$ through Eq. (1). In other words, the fracture strength $\sigma_c(a/P)$ for a specific precut specimen is simply proportional to inherent fracture strength $\sigma_{F(inh)}$ with the proportionality constant being a function of $a/P$.

*5.2 New expression for $G_{Ic}$*

It is worth noting that we have suggested Eq. (20) to replace Eq. (8). While Eq. (8) contains no information about cut characteristics, Eq. (2) does. Superficially, Eq. (8) seems to indicate a rather different origin of fracture from that revealed by Eq. (20): The fracture strength $\sigma_c$ of precut specimen would depend on $G_{Ic}$ according to Eq. (8) whereas we know from Eq. (20) that $\sigma_c$ explicitly depends on $\sigma_{F(inh)}$. Following the line of Eq. (8) at does $G_{Ic}$ of Eq. (6) depend on since we know it is not given by surface fracture energy? According to Eq. (6) and Eq. (18), $G_{Ic}$ is determined by inherent fracture strength $\sigma_{F(inh)}$ and P

$$G_{Ic} = w_F(4\pi P), \text{ with } w_F = (\sigma_{F(inh)})^2/2E \quad \text{and for } a \gg P, \quad (21)$$

where the work density $w_F$ is defined at the stress level of $\sigma_{F(inh)}$. If we take $\sigma_{F(inh)} = \sigma_b$, then insertion of $w_F$ in place of $w_c$ in Eq. (9b) produces $L_{fc} = 4\pi P$ by identification between Eq. (21) and Eq. (9a-b). Thus, the fractocohesive length has acquired geometric meaning, i.e., the meaning of P, as further discussed in Section 5.6.
We note that Eq. (21) is identical in form to the well-known expression[61, 62] of Thomas, proposed based on consideration of fracture of elastomers, a topic[35, 36] to be dealt with in our subsequent publication. In closing of Section 5.2, it is necessary to clarify what $G_{Ic}$ is and it is not: $G_{Ic}$ simply accounts for how much energy release is actually involved during fracture, and its magnitude is dictated by $w_F$ and P, as $K_{Ic}$ is by $\sigma_{F(inh)}$ and P shown in Eq. (18).

*5.3 Tip blunting, calculation of $G_{Ic}$ from stress saturation (SS)*

Because of the mathematical stress singularity, Irwin and others had to propose a plastic zone as depicted in Fig. 1a, leading to Eq. (4) and (5). However, the scenario also makes impossible to prescribe $K_{Ic}$: there is no account for what $r_p^*$ in the operational definition of Eq. (5) should be. This inherent difficulty stems from the perceived stress singularity. Fracture mechanics has identified an effective way to deal with this situation: it turns to the energy balance argument of Griffith and suggests that we can try to estimate $G_c$ even though $G_c$ is found to have little to do with surface fracture energy $\Gamma$. Existing theoretical attempts have been made by adhering to the literal meaning of $G_{Ic}$ as energy change per unit fracture surface area during fracture. Specifically, both Lake-Thomas model[63] for elastomers and Peppas model[43] for brittle plastics treat $G_{Ic}$ as a quantity involving fracture surface rather than fracture volume, i.e., $G_{Ic}$ is of the following general form

$$G_{Ic} = \psi_{LBS}(n_{LBS}E_b), \quad (22)$$

where $\psi_{LBS}$ is the average areal density of load-bearing strand (LBS) across fracture plane, $n_{LBS}$ is the average number of bonds per LBS, and $E_b$ is the bond dissociation energy. Since $G_{Ic}$ in such an expression is proportional to LBS areal density, it scales linearly stress in contrast to Eq. (6a) and Eq. (21) that involve an energy density, quadratic in stress. As characteristic of fracture mechanical account of fracture in terms of energy balance argument, the formulation of $G_c$ involves no information of local stress field at the tip and thus makes no reference to the geometric characteristic of the cut tip.

In practice, determination of $G_c$ is based on precut specimens involving introduction of intentional cut to a given material. Sharpness of precut tip depends on the material response to the protocol used to make the cut. In other words, through damage mechanics, material response produces certain characteristic tip bluntness, which may simplistically be depicted as effective radius of curve, P. We can expect P to vary greatly from silica glasses to metals and polymers although there appears to be little understanding of this scale in terms of the material physics. Existence of finite tip bluntness assures that stress buildup would cease[58, 59, 64, 65] at $P = r_{pl}$ as shown in Fig. 1c. In other words, we deal with the scenario of Fig. 1c instead of Fig. 1a, i.e., existence of stress saturation zone at cut tip independent of load level, and $K_{Ic}$, as shown in Eq. (18), acquires new meaning, given by inherent material strength and tip bluntness, contrasting its operational definition in Eq. (7). Since natural tip blunting varies in a narrow range, $K_{Ic}$ changes in a small range, i.e., $K_{Ic}$ (as well as $G_{Ic}$) appears constant for a given type of material.

Given the stress field description in Fig. 1c where $r_{pl} = P$, we can propose a picture to evaluate $G_{Ic}$ involved in the fracture. Since the stress in the SS zone reaches $\sigma_{F(inh)}$ at fracture, we estimate $G_c$ according to Fig. 17. In other words, we suggest that $G_c$ involves a volume given by ($\delta P t$) in the energy calculation. Upon crack propagation, the energy associated with the (yellow color online) failure region is given by $W_P = w_F(\delta P t)$, so that we have

$$G_{Ic}(\delta t) = w_F(\delta t P) \text{ or } G_{Ic} = w_F P \quad (23a)$$

where $w_F$ is given by

$$w_F = [\sigma_{F(inh)}]^2/2E \sim w_c, \quad (23b)$$

where $w_c$ given by Eq. (9b). In other words, based on Fig. 1c, as reveal by Fig. 7b, we have derived an expression for $G_{Ic}$, given by Eq. (23a-b). We can estimate the magnitude of $G_{Ic}$ once the tip bluntness P is known, along the basic mechanical characteristic: $\sigma_b$. Apart from a numerical prefactor, Eq. (23a-b) is Eq. (21). Moreover, Eq. (23a) is identical in form to the well-known expression[61, 62] of Thomas, proposed based on consideration of fracture of elastomers, a topic[35, 36] to be dealt with in our subsequent publication. While the picture in Fig. 17 is motivated by the observation of SS zone at the crack tip, there is some experimental evidence in the cases of hydrogels[66] and elastomers[67, 68]. Bases on Fig. 17 and Eq. (23a-b), we suggest that crack propagation does not require $G_{Ic}$; rather it involves $G_{Ic}$.

*5.4 Variation in $K_{Ic}$ due to change of P*

In the preceding section we have clarified the significance of $K_{Ic}$ in terms of Eq. (18). If one insists on using $G_{Ic}$, then Eq. (6) prescribes its magnitude through $K_{Ic}$. In other words, both $G_{Ic}$ and $K_{Ic}$ for a precut specimen trace their meanings to the fact that their magnitudes are determined by $\sigma_{F(inh)}$ and P. Separately, the meaning of P affords us a simple explanation for the spreads seen in Fig. 7e and 7f. For example, the origin of scattering in Fig. 7e presumably largely resides in the variation of the cut tip characteristics that result in different values for P, as proposed in the preceding subsection. Similarly, the variation of $K_{Ic}$ from 0.8 to 1.4 MPa.m$^{1/2}$ in Fig. 7f may dominantly have arisen from the difficulty to have the same effective tip bluntness in the precut specimens, i.e., from the variation of P by a factor of 3. In other words, the fracture condition, characterized by $K_{Ic}$, depends on the cut characteristics: sharper cut (lower value of P) produces lower $K_{Ic}$. This statement is consistent with the reported correlation of impact strength with the tip radius.[69] Having clarified the P dependence of $K_{Ic}$, it is clear that the stress plateau width depicted in Fig. 1c in terms of $r_{pl}$ is actually determined by the material's mechanical response and tip bluntness, i.e., $r_{pl}$ = P.

Since the spatial resolution $r_b$ of our birefringence observation is smaller than P, we are able to determine $\sigma_{F(inh)}$ from $\sigma_{tip}$ at fracture and show it to be comparable to $\sigma_b$. Given that $\sigma_{F(inh)} \sim \sigma_b$, we can contemplate a hypothetical case of arbitrarily sharp cut, e.g., with P approaching $l_{LBS}$, prepared not through mechanical forcing. Let us consider any appreciable value of $a$, e.g., 1 mm, our birefringence would merely report fracture at a tip stress at $r_b$ equal to $\sigma_{F(inh)}(2l_{LBS}/r_b)^{1/2}$ = 2.7 MPa at a critical load $\sigma_c$ = $\sigma_{F(inh)}(2l_{LBS}/a)^{1/2}$ = 0.19 MPa, given $l_{LBS}$ = 5 nm, $r_b$ = 5 μm and $\sigma_{F(inh)} \sim$ 60 MPa. Inability of PMMA to show molecularly sharp cut becomes an advantage for our birefringence observations. In other words, to be able to discern $\sigma_{tip}(P, K_{Ic})$ at fracture we need P > $r_b$ so that the inherent strength can be determined based on the limited spatial resolution of our birefringence observations.

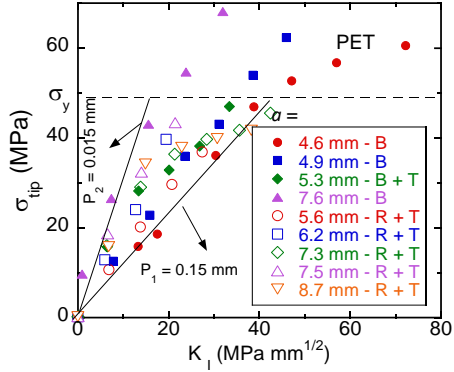

**Fig. 18.** Stress observed at the notch tip of PET specimens ($L_0 \times W_0 \times D_0$ = 100 × 20-25 × 0.25 mm$^3$ drawn at $V/L_0$ = 0.01 min$^{-1}$) as a function of $K_I$ for B- and R-type cuts. Specimens with "+ T" were thermally annealed at 80 °C to remove residual birefringence that was introduced during cutting. ● "4.6 mm – B" specimen is reproduced from Fig. 8. ▲ "7.6 mm – B" specimen was cut in liquid nitrogen, whereas ■ "4.9mm – B" and ♦ "7.6mm – B" specimens were cut at –20 °C.

Moreover, before yielding, as expected, the stress intensification in PET follows a similar trend as found in PMMA, e.g., showing an approximate linear relation between the local stress $\sigma_{tip}$ at the cut tip and load parameter $K_I$, as shown in Fig. 12a, for all three polymers. Here we applied different methods to make intentional pre-through-cuts. This has induced widely varying material responses, leading to different values for P. For example, for the same PET, depending on how the cut is made, we have different $K_I$ dependence of the tip stress, as shown in Fig. 18, corresponding to P having a range from 0.015 to 0.15 mm. We can examine the stress buildup at cut tip using birefringence measurements to compare with the conclusion made from Fig. 18. Figs. 19a-b shows that the stress plateau occurs at different distances from the tip, showing a similar qualitative trend as revealed by Fig. 18. For example, Fig. 19a shows the cessation of stress building around 0.12 mm whereas the local stress keeps climbing past 0.1 mm until 0.03 mm in Fig. 19b. Correspondingly, the normalized stress plateau, represented by the horizontal lines in Fig. 19a-b increases as the plateau width narrows, i.e., as P decreases.

### 5.5 Ductile cases: deviation from Dugdale model

Ductile PET and PC also depart strongly from the LEFM description of stress intensification, e.g., Fig. 1a. For example, the stress at cut tip ($\sigma_{tip}$) increases with the loading instead of exhibiting a limiting value independent of the load level $\sigma_0$. Specifically, before yielding and plastic deformation, $\sigma_{tip}$ increases linearly with $\sigma_0$. Upon yielding, the buildup of $\sigma_{tip}$ slows down, revealing a rather clear elastic-yielding transition that is absent in brittle PMMA. The onset of yielding and irrecoverable deformation (emergence of damage zone) is marked by a local stress level comparable to yield stress $\sigma_y$ observed from uncut specimens, confirming $\sigma_{Y(inh)} \sim \sigma_y$. This observation indicates that no intrinsic flaws are present to lower $\sigma_y$ relative to $\sigma_{Y(inh)}$. Since the local stress ceases to build up around cut tip at P, yielding does not occur below a certain load level, contrary to the classic Dugdale model prediction. In other words, the formation of the yielding zone occurs only when $\sigma_0$ is beyond a threshold as shown in Fig. 20. It is clear that no yielding zone emerges of a finite size $s$ until the far-field stress is 15% of the yield stress $\sigma_y$ for both PET and PC, in agreement with the yield characteristics revealed in Figs. 8 and Fig. 10. Since our experiments involve SEN whereas Dugdale model[70] treats the case of embedded through-cut, the Dugdale curve in Fig. 20 only serves to provide a reference. Dugdale curve passes through the origin, implying yielding at vanishingly low load because Dugdale model predicates stress divergence at the crack tip.

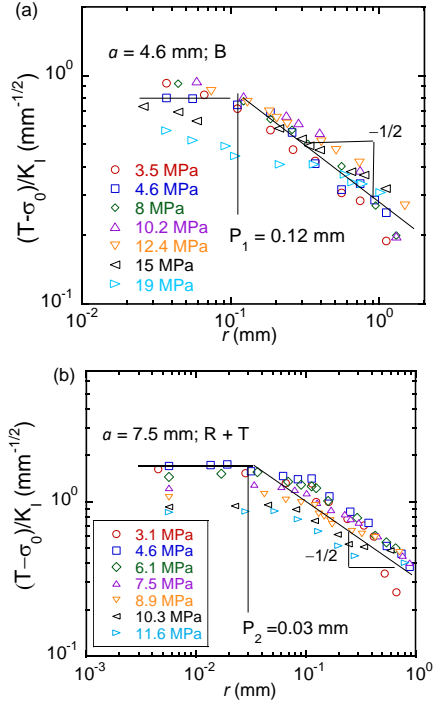

**Fig. 19.** Local tensile stress T as a function of distance $r$ from the notch tip for two precut PET specimens (a)-(b) whose mechanical responses during drawing have been described in Fig. 18, denoted respectively by △ and ●. The two different values of P depict respectively the spatial extent of the stress plateau zones, marked by $r_{pl}$ = P in Fig. 1c.

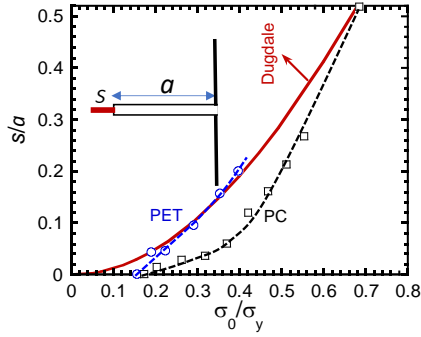

**Fig. 20.** Evolution of the damaged zone in front of the crack tip as a function of the ratio of applied nominal load $\sigma_0$ to the yield stress $\sigma_y$. The red (color online) solid line represents the Dugdale model: $s/a = \sin^2(\pi\sigma_0/4\sigma_y)$. The PET specimen is from Fig. 8 and the PC specimen is from Fig. 10, where the dashed lines are drawn through the data points to show the trends. Yielding is not observed until applied load is sufficiently high to cause localized yielding at the crack tip, at odds with onset of yielding at arbitrarily low value of the stress ratio $\sigma_0/\sigma_y$.

*5.6 Fractocohesive length*

The clarification of the meaning of P as related to $K_{Ic}$ in Eq. (18) allows us to further discuss the significance of $L_{fc}$ in Eq. (9a), e.g., what factors affect its magnitude or what meaning it has. Inserting Eq. (2a) into Eq. (9a) through Eq. (6), we have, in the limit of linear elasticity,

$L_{fc} = 2\pi a^*$,        (24a)

where use is made of Eq. (9b). Separately, insertion of Eq. (21) into its definition Eq. (9a-b) produces another expression for $L_{fc}$ as

$L_{fc} = 4\pi P(w_F/w_c) = 4\pi P(\sigma_{F(inh)}/\sigma_b)^2 \sim 4\pi P$.        (24b)

Thus, $L_{fc}$, $a^*(\approx 2P)$ and P all have the same origin: natural tip bluntness. For the present glassy polymers, $L_{fc}$ is on the order of 1 mm. For window glasses, the cut would be so sharp that any crack only needs to be very tiny to weaken them, i.e., P is 1 nm for silica glasses instead of the present P = 0.1 mm for glassy polymers. Like $K_{Ic}$, P and $L_{fc}$ may be regarded as a material parameter. Indeed, the relationship between $L_{fc}$ and P can be used to reveal the true meaning for $G_{Ic}$. In other words, since $w_c$ of Eq. (9b) is a material constant, $L_{fc}$ of Eq. (9a) is another measure of $G_I$. This discussion concerning the meaning of $L_{fc}$ appears consistent with that[48] in a previous study, which suggested $L_{cf}$ to be related to $a^*$, consistent with Eq. (24a). We note in passing, however, since the meaning of P is illustrated in Fig. 1c, i.e., P = $r_{pl}$, and since $L_{fc}$ is one order of magnitude greater than P according to Eq. (24b), it is perhaps more accurate to refer to P rather than $L_{fc}$ when discussing the size of the stress saturation (SS) zone.

Given the interpretation of P in Fig. 17 as a parameter to indicate the extent of structural failure, tip bluntness and the size of SS zone, the trend of the data in Fig. 12b corresponds to P decreasing during elastic drawing and effectively increasing upon crossing the dividing line (denoting the elastic-yielding transition) to involve massive plastic deformation. As noted, the change is considerably higher for PET and PC than for PMMA. It is reasonable that characteristics of the cut tip respond to the loading level. In other words, there is non-negligible change in the geometry of the opening associated with the cut that amounts to the tip getting sharper. The various cutting methods typically generate an incomplete through cut, and at an increased load, the cut becomes closer to an ideal through-cut. The degree to which this response takes place depends on the nature of the cut and the material. PMMA shows a weak response. In short, the reality is far more complex than what can be captured by the ideal analysis of Eq. (3a). In contrast to the behavior shown in Fig. 12b, it is worth mentioning that the $\Omega$ values at the poles of circular holes, as a measure of the stress intensification, are high even at the beginning of drawing, as shown in Figs. 13b and 14b. Here the geometric characteristic of circular holes can be expected to change little during extension.

**6. Summary**

In this work we show based on three commodity glassy polymers that their fracture and failure behaviors can be understood in terms of the concept of stress intensification (SI), familiar from fracture mechanics. Using birefringence and the stress-optical rule to correlate birefringence to stress, we describe spatial distribution of the stress field at cut tip. We find that the inherent fracture strength $\sigma_{F(inh)}$ and yield strength $\sigma_{Y(inh)}$, estimated by the local stress at cut tip during fracture and yielding, are respectively comparable to the breaking stress $\sigma_b$ and yield stress $\sigma_y$ measured from uncut specimens. This finding implies that the observed mechanical strength in cut-free samples reflects the true strength of such polymers albeit there is a numerical difference between plane stress and plane strain states. This occurs for present glassy polymers because they exhibit a rather sizeable SS zone, on the order of P = 0.1 mm. Unless these polymers indeed contain inclusions (to be regarded as flaws) of size greater than 0.1 mm, which is generally not the case, understanding of their fracture behavior in absence of intentional precut requires us to go beyond[14,71] the classical domain of fracture mechanics. On the other hand, in presence of sizeable cut, e.g., with $a \gg P$, the stress intensification analysis of fracture mechanics provides the framework for brittle polymers. For the present ductile PET and PC, post-tip-yielding is not straightforward to analyze. Without building in constitutive information such as strain hardening behavior (due to chain network tightening), Dugdale model cannot be used to describe the localized necking that starts at the tip and stabilizes during advancement of the necking across the specimen in the case of PET and gives in to crack propagation in the case of PC.

For glassy polymers, a typical through-cut introduces a pertinent length scale P (~ 0.1 mm) resolvable within the reach of a standard 4K video from a 4K CCD camera connected to a microscope-objective lens that affords us a spatial resolution at ca. 1 μm. The bluntness of the precut determines the value of P. We found using birefringence measurements that the stress buildup at cut tip, denoted by $\sigma_{tip}$, ceases to increase within a distance P from cut tip and stays below the inherent strength till fracture. Specifically, $\sigma_{tip}$ is found to grow linearly with the stress intensity factor $K_I$, with a proportionality constant on the order of $(2\pi P)^{1/2}$. A specimen with precut of sufficient size is mechanically weaker than an uncut specimen of the same polymer, due largely to the well-known SI. The degree of SI depends on how sharp the precut is made to be and therefore on the material response, i.e., on $a/P$, where for the glassy polymers, the sharpness seems to be upper bounded by P, regarding the cut as conforming to a cylinder of effective diameter 2P. In terms of this characteristic length scale P, below the critical load for fracture, the observed stress plateau (or SS) at P has little in common with the concept of plastic zone.

The birefringence measurements of local stress field allow us to show that inherent strengths $\sigma_{F(inh)}$ and $\sigma_{Y(inh)}$ are respectively comparable to breaking and yield stresses $\sigma_b$ and $\sigma_y$ that characterizes mechanical responses of uncut specimens. Consequently, we can identify the origin of toughness expressed by the critical stress intensity factor $K_{Ic}$. In other words, we can see, perhaps for the first time, what material properties determines the magnitude of $K_{Ic}$: Eq. (18) shows that the value of $K_{Ic}$ can be predicted from $\sigma_b \sim \sigma_{F(inh)}$, apart from knowing the value of P, for which there is an illustration in Fig. 17 but there is currently no theory. Given Eq. (18), a strong material with higher $\sigma_{F(inh)}$ can be expected to have higher toughness, i.e., higher $K_{Ic}$. Moreover, any mechanism to cause material respond with greater P will also enhance toughness. Critical stress intensity factor $K_{Iy}$ for yielding at crack tip can be similarly estimated from $\sigma_y \sim \sigma_{Y(inh)}$ and P, following an expression analogous to Eq. (18).

The critical energy release rate $G_{Ic}$ for brittle fracture can now be used beyond its operational definition of Eq. (6), through the combination of Eqs. (6) and (18), i.e., from $K_{Ic}$. For example, the magnitude of $G_{Ic}$ is determined by the work of fracture (involving $\sigma_b \sim \sigma_{F(inh)}$) and P, as shown in Eq. (21), as $K_{Ic}$ is determined by $\sigma_{F(inh)}$ and P in Eq. (18). In absence of a theory to derive P from material's mechanics characteristics, it is intractable to have a theoretical derivation of $G_{Ic}$ based on microscopic modeling. While $K_{Ic}$ in Eq. (18) or $G_{Ic}$ in Eq. (21) is a material parameter useful to rank the resistance of crack propagation, it is $\sigma_{F(inh)}$ that characterizes the mechanical strength of uncut mechanics for glassy polymers. An increase in P has a favorable effect on $K_{Ic}$ or $G_{Ic}$. Thus, we expect Eqs. (18) and (21) to have provide working strategies to toughen polymers.

The present study has permitted us to explore the difference between the theoretical description and experimental reality regarding fracture behaviors of various glassy polymers. Specifically, our data reveals how the inherent fracture strength in combination with a characteristic length scale P determine the toughness. Here P has essentially lent meaning for the fractocohesive length $L_{fc}$ in Eq. (9a). $L_{fc}$ is no longer elusive since $G_{Ic}$ is clear from Eq. (21). The meaning of $L_{fc}$ was unclear when $G_{Ic}$ has only its operational definition in Eq. (6). In other words, Eq. (9a) should take the form of Eq. (21) to reveal the causality: $G_{Ic}$ is dictated by the length scale P. In other words, $G_{Ic}/w_c$ produces a length scale in Eq. (9a) because $G_{Ic}$ involves a characteristic length scale P as shown in Eq. (21). For glassy polymers, P appears to be on the order of 0.1 mm or so. This prescribes toughness $K_{Ic}$ to be of a magnitude given by a product of $\sigma_b \sim \sigma_{F(inh)}$ = 50 MPa and $(2\pi P)^{1/2}$, i.e., $K_{Ic} \sim 1$ MPa.m$^{1/2}$. Such an account does not run into any difficulty that was previously encountered when the critical Griffith-Irwin energy release rate $G_{Ic}$ was found to be several orders of magnitude higher than the surface fracture energy: $G_{Ic}$ merely shows what the fracture energetically involves and describes an effect not the cause. Separately, because of the sizable P, mechanical characteristics of uncut glassy polymers such as brittle fracture can be described without reference to LEFM: fracture of uncut specimens initializes when force imbalance occurs[14] between chain tension and topological static friction that has prevented chain pullout before fracture.

This work is supported, in part, by the Polymers program at the National Science Foundation (DMR-1905870) that also provided REU supplemental funding for CC's summer internship. SQW acknowledges many email exchanges with Zhigang Suo, from whom he learned the essence of fracture mechanics. We also acknowledge participation of Mr. Zehao Fan in the discussion of the birefringence measurements and finally thank Dr. X. Yu in Auriga Polymers Inc. for providing us the PET sheets.

*Credit author statement*

**Travis Smith**: Development of quantitative birefringence measurements, Data curation on PET and PC, Formal Analysis, Validation, Writing - reviewing and editing. **Chaitanya Gupta**: Development of quantitative birefringence measurements, Data curation on PMMA, Formal Analysis, Validation, Writing - reviewing and editing. **Caleb Carr**: Development of quantitative birefringence measurements. **Shi-Qing Wang**: Supervision, Conceptualization, Methodology, Formal Analysis, Validation, Writing - original draft, Writing - reviewing and editing.